\documentclass[preprint,12pt]{elsarticle}
\pdfoutput=1

\usepackage{despaper}
\usepackage{fontawesome}

\journal{Nonlinear Analysis: Hybrid Systems}

\begin{document}

\begin{frontmatter}



\title{Usability Aware Secret Protection with Minimum Cost}


\author[1]{Shoma Matsui}
\ead{s.matsui@queensu.ca}

\author[2]{Kai Cai\corref{cai}}
\ead{kai.cai@eng.osaka-cu.ac.jp}

\address[1]{Department of Electrical and Computer Engineering, Queen's University, Kingston, Ontario K7L 3N6, Canada}
\address[2]{Department of Electrical and Information Engineering, Osaka City University, 3-3-138 Sugimoto, Sumiyoshi-ku, Osaka 558-8585, Japan}
\cortext[cai]{Corresponding author}

\begin{abstract}
In this paper we study a cybersecurity problem of protecting system's secrets with multiple protections and a required security level, while minimizing the associated cost due to implementation/maintenance of these protections as well as the affected system usability. The target system is modeled as a discrete-event system (DES) in which there are a subset of marker states denoting the  services/functions provided to regular users, a subset of secret states, and multiple subsets of protectable events with different security levels. We first introduce {\em usability-aware cost levels} for the protectable events, and then formulate the security problem as to ensure that every system trajectory that reaches a secret state contains a specified number of protectable events with at least a certain security level, and the highest usability-aware cost level of these events is minimum. We first provide a necessary and sufficient condition under which this security problem is solvable, and when this condition holds we propose an algorithm to solve the problem based on the supervisory control theory of DES. Moreover, we extend the problem to the case of heterogeneous secrets with different levels of importance, and develop an algorithm to solve this extended problem. Finally, we demonstrate the effectiveness of our solutions with a network security example.
\end{abstract}




\end{frontmatter}



\section{Introduction}\label{sec:introduction}

In real networked systems, risks and threats due to cybersecurity breach are increasingly prominent. Effectively protecting systems so that confidential information remains undisclosed to adversarial access has become an indispensable system design requirement \cite{P.Barrett2018, Brooks2018}. 

Recently cyber-physical systems (CPS) has emerged to be a general modeling framework for real networked systems consisting of both physical and computational components. 
CPS security issues have attracted much attention in the literature \cite{Hoffman:2009,Teixeira:2012,Modi:2013,Pasqualetti:2015}. For example, \cite{Teixeira:2012} discusses several attack scenarios with a typical architecture of networked control systems.

Focusing primarily on the abstracted level of dynamic systems, the research community of discrete-event systems (DES) has actively studied a number of security related problems. 
An ealier and widely investigated problem is {\em opacity} (e.g. \cite{Lin:2011aut,Hadjicostis:2011tase,Lafortune:2018arc,Toni:2017tacSO}). This is a system property under partial observation such that an intruder cannot infer a given set of {\em secrets} by (passively) observing the system behavior. Depending on the definitions of secrets, opacity takes different forms. Recent work extends opacity notions to networked, nondeterministic settings as well as Petri net models (e.g. \cite{Yin:2019,Xie:2020,Lan:2020}).

Another well studied problem is {\em fault-tolerance} and {\em attack-resilience} (e.g. \cite{Moor:2016,Fritz:2018,Lin:2019,Paape:2020,Yao:2020}). This is a design requirement that a supervisory controller should remain (reasonably) operational even after faults occur in the system or the system is undre malicious attacks.  

\emph{Intrusion detection} is another problem that has recently attracted much interest  (e.g. \cite{Lafortune:2018aut,Zhang:2018wodes,Agarwal:2019smcConf,Gao:2019smcConf,Goes:2020}). In this problem, the aim of the system administrator is to detect invasion of intruders by identifying abnormal behaviors in the system; if invasion is detected, an alarm can be set off before any catastrophic damage can be done by intruders.

From a distinct perspective, in our previous work a {\em minimum cost secret protection} problem is introduced \cite{KaiCai:2018Japan,KaiCai:2019cdc,Ma:2020,Matsui:2021}. This problem is concerned with the scenario that the system contains sensitive information or critical components to which attackers want to gain access, and attackers may be able to observe all events and disguise themselves as regular users without being detected. Then the system administrator is required to protect the sensitive information or critical components with proper security levels, while practically balance with the costs associated with the implementation and maintenance of the adopted protection methods.

In this paper, we make two important generalizations of the minimum cost secret protection problem. First, we take into account system's {\em usability}, which means regular users' convenience of using various services and funcitions provided by the system. These services and functions for regular users are often different from sensitive information or critical components that need to be protected. However, bad choices of protection points/locations may simultaneously affect access to services/functions by regular users.  
For example, when setting up a password to protect a user's credit card information, it is not reasonable that the user has to input the same password in order to access any websites or files. If system's usability is significantly reduced owing to setting up too many protections at inappropriate locations, users may stop using the system and this can be costly (to different extent depending on specific situations/applications). Accordingly, we formulate usability as another source of protection cost, in addition to the implementation/maintenance cost of protection methods (considered in previous work).

The second extension to the minimum cost secret protection problem is that on top of the usability consideration, we further differentiate sensitive information and critical components (or simply secrets) with distinct degrees of importance. This is a typical situation in practice; for instance, in e-commerce, customers' email addresses and credit card numbers are both sensitive information, but it is common that the latter are deemed more important and expected to be protected with stronger measures. 
Accordingly, we formulate heterogeneous secrets by a partition on the set of all secrets, and require that more important secrets be protected using more secure methods (while system usability still needs to be balanced). 

The main contributions of this work are summarized as follows. 
\begin{itemize}
  \item A novel concept of system's usability is introduced and formulated. This notion was absent in our previous work \cite{KaiCai:2018Japan,KaiCai:2019cdc,Ma:2020,Matsui:2021}, and to our best knowledge is new in the DES security literature. Roughly speaking, the formulation of usability is based on counting the number of affected services/functions provided to regular users when a protection is implemented at a certain location, and comparing this number to a prescribed threshold to determine if such a protection is too costly.  
  \item A new usability-aware minimum cost secret protection problem is formulated, its solvability condition characterized, and an solution algorithm designed.  In constrat to the problem without usability consideration \cite{KaiCai:2018Japan,KaiCai:2019cdc,Ma:2020,Matsui:2021}, in our problem less secure protection methods that significantly undermine usability may be just as costly as more secure methods that make little impact on usability. This new feature due to usability makes our problem more challenging because security levels and cost levels of the same protection methods are generally different, and hence need to be treated separately (security levels and cost levels are treated as the same in \cite{KaiCai:2018Japan,KaiCai:2019cdc,Ma:2020,Matsui:2021} since usability is not considered).
  \item A new minimum cost secret protection problem featuring both usability awareness and heterogeneous secrets is formulated, its solvability condition characterized, and an solution algorithm developed. Not only are the formulated problem and developed solution algorithm new as compared to the existing literature, but also this problem covers a general and practical scenario in the context of secret protection.
\end{itemize}

The rest of this paper is organized as follows. Section~2 introduces system model and definitions of cost; Section~3 formulates two usability aware minimum cost secret protection problems; Section~4 solves the first problem in which all secrets are deemed equally important, while Section~5 solves the second problem in which the secrets have different importance; finally in Section~6 we state our conclusions and future work.

\section{System Model}\label{sec:model}

Consider that a system administrator needs to protect all secret information in the system. The administrator desires to do so in such a way that every secret is protected with at least a certain number of protections and these protections are of at least a certain security level. 

Meanwhile, the administrator needs to balance secret protection with the associated cost. There are two sources of cost often considered in practice. One is the cost of purchasing, implementing, and maintaining the device or program for protection. This cost evidently varies depending on the means of protection; for example, a biometric device is much more costly than a password protection. Correspondingly, the higher the cost is, the higher the {\em security level} of the protection becomes.

The other source of cost is due to that secret protection can have the side effect of negatively impacting the convenience of regular users of the system. Unlike intruders, regular users when using the system do not always try to see the secret information (e.g. personal data), but more often use various services that the system provides (e.g. watching a movie, reading an e-book, launch an app). If protecting secrets simultaneously requires regular users to undergo many security checks before using any services, user experience or system's {\em usability} will decline, and if this causes users to stop using the system, the cost can be significant.  

In this section, we will formulate the above-described system and cost considerations for secret protection. Our objective is to design for the administrator a {\em protection policy} that ensures the required level of secret protection while minimizes the incurred cost.


To model the system, we employ the framework of discrete-event systems (DES) \cite{wonham2019supervisory,cassandras2009introduction}, and consider the system modeled as a finite-state automaton
\begin{equation}\label{eq:plant:model}
    \mathbf{G} = (Q, \Sigma, \delta, q_0, Q_m).
\end{equation}
Here $Q$ is the set of states, $\Sigma$ the set of events, $\delta: Q
\times \Sigma \to Q$ the (partial) transition function,\footnote{It is sometimes convenient to view $\delta$ as a set of triples: $\delta = \{(q,\sigma,q') \mid (q,\sigma) \mapsto q'\}$.} $q_0 \in Q$ the initial state, and $Q_m \subseteq Q$ the set of {\em marker states} which models the set of services/functions provided by the system to its users.
We denote by $Q_s \subseteq Q$ the set of {\em secret states} in
$\mathbf{G}$; no particular relation is assumed between $Q_s$ and $Q_m$, i.e. a secret state may or may not coincide with a marker state.
In addition we extend the transition function $\delta$ to $\delta: Q \times \Sigma^* \to Q$ (where $\Sigma^*$ is the set of all finite-length strings of events in $\Sigma$ including the empty string $\epsilon$) in the
standard manner, and write $\delta(q, s)!$ to mean that string $s$ is defined at state $q$. The {\it closed behavior} of {\bf G}, written $L({\bf G})$, is the set of all strings that are defined at the initial state $q_0$:
\begin{align*}
 L({\bf G}) = \{s \in \Sigma^* \mid \delta(q_0, s)!\}.   
\end{align*}
Also define the {\it marked behavior} of {\bf G}:
\begin{align*}
 L_m({\bf G}) = \{s \in L(\bf G) \mid \delta(q_0, s)! \ \&\ \delta(q_0, s) \in Q_m\}.   
\end{align*}
That is, every string in $L_m({\bf G})$ is a member of the closed behavior $L({\bf G})$, and moreover reaches a marker state in $Q_m$.

A state $q \in Q$ is {\em reachable} (from the initial state $q_0$) if there is a string $s$ such that $\delta(q_0, s)!$ and $\delta(q_0, s)=q$. 
A state $q \in Q$ is {\em co-reachable} (to the set of marker states $Q_m$) if there is a string $s$ such that $\delta(q, s)!$ and $\delta(q, s) \in Q_m$. $\mathbf{G}$ is said to be {\em trim} if every state is both reachable and co-reachable. Unless otherwise specified, we consider trim automaton {\bf G} for the system model in the sequel.

In practice, not all events in the system can be protected by the administrator for reasons such as exceeding administrative permissions. Thus we partition the event set $\Sigma$ into a disjoint union of the subset of {\em protectable}
events $\Sigma_p$ and the subset of {\em unprotectable} events $\Sigma_{up}$, namely
$\Sigma = \Sigma_p \disjoint \Sigma_{up}$. 
Moreover, protecting different events in $\Sigma_p$ may incur different costs. As described at the beginning of this section, we consider two sources of cost. 

For the first source of purchasing/implementing/maintaining the protection device/program, we partition the set of protectable events $\Sigma_p$ further into $n$ disjoint subsets $\Sigma_i$ where
$i \in \{0, 1, \dots, n-1\}$, namely
\begin{equation} \label{eq:Sigmai}
    \Sigma_p = \bigdisjoint_{i=0}^{n-1} \Sigma_i.
\end{equation}
The index $i$ of $\Sigma_i$ indicates the cost level 
when the system administrator protects one or more events in $\Sigma_i$; the larger the index $i$, the higher the cost level of protecting events in $\Sigma_i$. For simplicity we assume that the index is the deciding factor for the first source of cost; that is, the cost of protecting one event in $\Sigma_i$ is sufficiently higher
than the cost of protecting all events in $\Sigma_{i-1}$. While this assumption might be restrictive, it is also reasonable in many situations: for example, the cost of purchasing/installing/maintaining a biometric sensor is more costly than setting multiple password protections. 
Since this source of cost is directly related to the strength of protection, we will also refer to these cost levels as {\em security levels}.

For the second source of cost regarding regular users' convenience, we investigate the impact of protecting an event $\sigma \in \Sigma_p$ at a state $q$ on the {\em usability} of services/functions provided by the system (which are modeled by the marker states in $Q_m$). In particular, we define for each pair $(q,\sigma)$, with $\delta(q,\sigma)!$, the following set of non-secret marker states that can be reached from the state $\delta(q,\sigma)$: 
\begin{align} \label{eq:U}
    U(q,\sigma) := \{ q' \in Q_m \setminus Q_s \mid (\exists s \in \Sigma^*) \delta(\delta(q,\sigma),s)! \ \&\  \delta(q,\sigma s) = q'\}.
\end{align}
This $U(q,\sigma)$ is the set of (non-secret) marker states that would be affected if $\sigma$ is protected at $q$; namely, regular users would also have to go through the protected $\sigma$ in order to use any of the services in this set. The reason why we focus on marker states that are {\em not} secrets is because it is unavoidable to cause inconvenience of the users if the services/functions to be used coincide with the secrets to be protected.

With the set defined in (\ref{eq:U}), it is intuitive that the cost of protecting $\sigma$ at $q$ is large (resp. small) if the size of this set, i.e. $|U(q,\sigma)|$, is large (resp. small). In case the cost is overly large, this event $\sigma$ (at $q$) belonging to (say) $\Sigma_i$ (i.e. the $i$th cost level of the first source) may be just as costly as those events in one-level higher $\Sigma_{i+1}$. For example, if setting up a password at a particular point to protect a secret simultaneously requires all regular users to enter a password for most services the system provides, this could largely reduce the users' satisfaction; hence this password protection may be as costly as using a biometric sensor (when the latter is used to protect a secret but affecting no regular users' experience).

As for how large this cost (measured by $|U(q,\sigma)|$) should $\sigma$ at $q$ be treated as having one-level higher cost is case dependent: different systems (or business) have different criteria. Thus we consider using a positive integer $T (\geq 1)$ as a threshold number: if the cost of the second source exceeds this threshould, i.e. $|U(q,\sigma)| \geq T$, the event $\sigma$ at $q$ belong to $\Sigma_i$ (say) will be treated as having the same cost level as those in $\Sigma_{i+1}$.
The more important the system deems user experience, the smaller threshold $T$ should be set. As a final note, the same event $\sigma$ at different $q$ generally has different $|U(q,\sigma)|$; hence this second souce of cost is state-dependent (in contrast with the state-independent first source of cost). 

With the above preparation, we now synergize the aforementioned two souces of cost as follows. Consider the partition of $\Sigma_p$ in (\ref{eq:Sigmai}) and let $T \geq 1$ be the threshold. First define 
\begin{align} \label{eq:C0}
    C_0 := \{ (\sigma, |U(q,\sigma)|) \mid q \in Q \ \&\ \sigma \in \Sigma_0 \ \&\ \delta(q, \sigma)! \ \&\ |U(q,\sigma)|<T \}.
\end{align}
Thus $C_0$ is the set of pairs in which the event belongs to $\Sigma_0$ (the lowest level of the first cost) and the $|U(q,\sigma)|$ (the second cost) is below the threshold $T$. In other words, these events at their respective states are the least costly ones when the first and second costs combined. 

Next for each $i \in \{1,\ldots,n-1\}$, define 
\begin{align} \label{eq:Ci}
    C_i :=& \{ (\sigma, |U(q,\sigma)|) \mid q \in Q \ \&\ \sigma \in \Sigma_i \ \&\ \delta(q, \sigma)! \ \&\ |U(q,\sigma)|<T \} \notag\\
    &\cup \{ (\sigma, |U(q,\sigma)|) \mid q \in Q \ \&\ \sigma \in \Sigma_{i-1} \ \&\ \delta(q, \sigma)! \ \&\ |U(q,\sigma)| \geq T \}.
\end{align}
As defined, $C_i$ is the union of two sets of pairs. The first set is analogous to $C_0$ (here for events in $\Sigma_i$). The second set is the collection of those pairs in which the event belongs to $\Sigma_{i-1}$ (one lower level of the first cost) and the $|U(q,\sigma)|$ (the second cost) is larger than or equal to the threshold $T$. 
Thus the events corresponding to the second set have different levels when only the first cost is considered and when the two costs are combined.

Finally define 
\begin{align} \label{eq:Cn}
    C_n := \{ (\sigma, |U(q,\sigma)|) \mid q \in Q \ \&\ \sigma \in \Sigma_{n-1} \ \&\ \delta(q, \sigma)! \ \&\ |U(q,\sigma)| \geq T \}.
\end{align}
Thus $C_n$ is the set of pairs in which the event belongs to $\Sigma_{n-1}$ (the highest level of the first cost) and the $|U(q,\sigma)|$ (the second cost) exceeds the threshold $T$. That is, these events at their respective states are the most costly ones when the first and second costs combined. 

It is convenient to define the set of events corresponding to $C_i$ ($i \in [0,n]$), by projecting the elements (i.e. pairs) to their first components. Hence for $i \in [0,n]$ we write
\begin{align} \label{eq:SigmaCi}
    \Sigma(C_i) := \{\sigma \mid (\exists q \in Q) (\sigma, |U(q,\sigma)|) \in C_i\}.
\end{align}
From (\ref{eq:C0})-(\ref{eq:Cn}), it is evident that 
\begin{align} \label{eq:SigmaC}
    \Sigma(C_0) \subseteq \Sigma_0,\quad \Sigma(C_{n}) \subseteq \Sigma_{n-1},\quad (\forall i \in [1,n-1]) \Sigma(C_i) \subseteq \Sigma_{i-1} \cup \Sigma_i.
\end{align}

To illustrate the system modeling and cost definitions presented so far, we provide the following example, which will also be used as the running example in subsequent sections.

\begin{exmp}\label{exmp:model}
    \begin{figure}[htp]
        \centering
        \begin{tikzpicture}[automata]
\node[state, initial] (0) {$q_0$};
\node[state, above right=of 0] (1) {$q_1$};
\node[state, right=of 1] (2) {$q_5$};
\node[state, below right=of 0] (3) {$q_2$};
\node[state, right=of 3] (4) {$q_6$};
\node[state, below right=of 2, secret1] (s2) {$q_8$};
\node[state, above=of s2, accepting, secret1] (s1) {$q_7$};
\node[state, below=of s2] (5) {$q_9$};
\node[state, right=of 5, accepting, secret1] (s3) {$q_{10}$};
\node[state, below=of 3, accepting] (m1) {$q_3$};
\node[state, above=of 2, accepting] (m2) {$q_4$};

\path
(0) edge node{$\sigma_0$} (1)
(1) edge node{$\sigma_5$} (2)
(0) edge[swap] node{$\sigma_1$} (3)
(3) edge node{$\sigma_2$} (1)
(1) edge node{$\sigma_6$} (4)
(3) edge[swap] node{$\sigma_5$} (4)
(2) edge node[near end]{$\sigma_7$} (s1)
(2) edge[swap] node{$\sigma_8$} (s2)
(4) edge[swap] node{$\sigma_9$} (5)
(s1) edge node{$\sigma_8$} (s2)
(s2) edge node{$\sigma_9$} (5)
(5) edge node{$\sigma_{10}$} (s3)
(3) edge[bend left] node{$\sigma_3$} (m1)
(m1) edge[bend left] node{$\sigma_4$} (3)
(2) edge[bend left] node{$\sigma_3$} (m2)
(m2) edge[bend left] node{$\sigma_4$} (2)
;
\end{tikzpicture}
        \caption{System $\mathbf{G}$: initial state $q_0$ (circule with an incoming arrow), marker state set $Q_m=\{q_3, q_4, q_7, q_{10}\}$ (double circles), secret state set $Q_s = \{q_7, q_8, q_{10}\}$ (shaded circles)}\label{fig:exmp:model:plant}
    \end{figure}
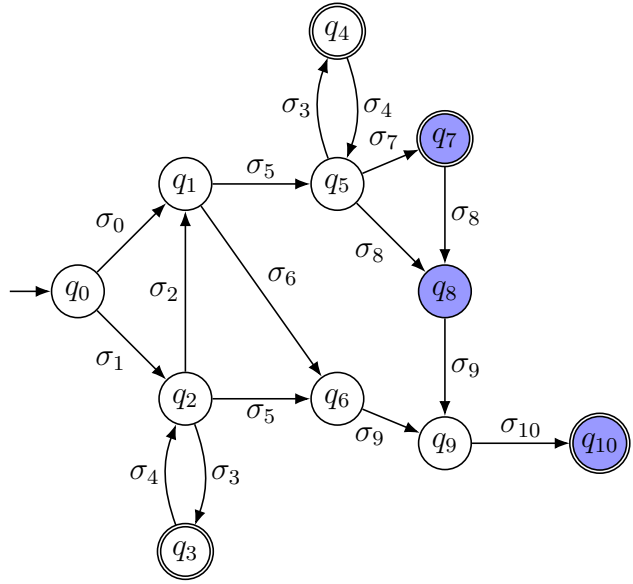
    The finite-state automaton $\mathbf{G}$ in \cref{fig:exmp:model:plant} represents a
    simplified system model of using a software application in which there are three
    restricted realms. There are also four services that this system provides. Consider that this application works according to the
    users' permission levels. Several authentication
    points can be (though need not be) set up so that the users have to pass them in order to obtain the permission to reach the restricted realms. 
    States $q_7$, $q_8$ and $q_{10}$ represent the restricted realms modeled as secret states, i.e. $Q_s = \{q_7, q_8, q_{10}\}$. On the other hand, states $q_3$, $q_4$, $q_7$, and $q_{10}$ represent the services provided by the system and hence the set of marker states is $Q_m=\{q_3, q_4, q_7, q_{10}\}$. 
    Thus $q_7$ and $q_{10}$ are simultaneously marker and secret states.

    The initial state $q_0$ indicates that a user is about to log into the
    system. Accordingly, events $\sigma_0$ and $\sigma_1$ represent logging
    into the system as a regular user or a system administrator
    respectively; then $q_1$ and $q_2$ mean that the user has logged in
    corresponding to $\sigma_0$ and $\sigma_1$ respectively. Typically, an
    administrator has higher-level permission in the system compared to a
    regular user. Also, $\sigma_2$ indicates switching permission from the
    administrator to a regular user, $\sigma_5$ denotes launching the application, and $\sigma_6$ means that a regular user launches the
    application with the administrative permission, e.g. \emph{sudo} in
    Unix-like operating systems. Events $\sigma_3$ and $\sigma_4$ are respectively the starting and finishing actions of using a system service. Moreover, $\sigma_7$ and $\sigma_8$ indicate the
    authentication points to obtain access to the secret states $q_7$ and $q_8$. On the other
    hand, the administrative realm denoted by the secret state $q_{10}$ requires users to pass
    two-factor authentication represented by $\sigma_9$ (first factor) and
    $\sigma_{10}$ (second factor). In order to keep secret states secure, the
    system administrator needs to configure several authentication points for
    restrict access.

    According to the above description, the set of protectable events is 
    \begin{align*}
        \Sigma_p = \{\sigma_0,\sigma_1,\sigma_5,\sigma_6,\sigma_7,\sigma_8,\sigma_9,\sigma_{10}\} 
    \end{align*}
    which can be partitioned into four different cost/security levels (low to high): 
    \begin{align} \label{eq:exSigmai}
    \Sigma_0 = \{\sigma_0,
    \sigma_1, \sigma_5\},\quad \Sigma_1 = \{\sigma_6, \sigma_7, \sigma_8\},\quad    \Sigma_2 = \{\sigma_9\},\quad
    \Sigma_3 = \{\sigma_{10}\}.
    \end{align}
    That is, $\Sigma_p = \Sigma_0 \disjoint \Sigma_1 \disjoint \Sigma_2 \disjoint \Sigma_3$ and $n = 4$. This is the first source of cost we consider, which corresponds to the level of security of these events. The remaining events are deemed unprotectable, i.e. $\Sigma_{up} = \{\sigma_2, \sigma_3,
    \sigma_4\}$. 
    
    For the second source of cost due to usability (user experience), in this example we set the threshold $T=2$, namely if protecting an event at a state affects two or more (non-secret) services provided by the system, this cost is deemed so large that the event at the state needs to be move one level up in terms of the total cost. In fact in ${\bf G}$, there are exactly two marker states that are not secret states: $q_3, q_4$; hence if both these two states are affected when protecting an event at a state, the threshold is reached. 
    
    Inspecting the set $U(q,\sigma)$ as defind in (\ref{eq:U}), we find $U(q_0,\sigma_1) = \{q_3,q_4\}$ because $\delta(q_0, \sigma_1 \sigma_3) = q_3$ and $\delta(q_0, \sigma_1 \sigma_2 \sigma_5 \sigma_3) = q_4$. As a result, $|U(q_0,\sigma_1)|=2=T$ and $\sigma_1 \in \Sigma_0$ at $q_0$ must be moved one level up in the total cost. Continuing this inspection, in fact $U(q_0,\sigma_1)$ is the only case where the threshold $T=2$ is reached. Also note that event $\sigma_5$ has different $|U(\cdot,\sigma_5)|$ at different states where it is defined: $|U(q_1,\sigma_5)|=1$ whereas $|U(q_2,\sigma_5)|=0$. This shows that the second cost is state-dependent. 
    
    Finally we present the cost level sets with the two sources of cost combined:
    \begin{align}
        C_0 &=\{(\sigma_0, 1), (\sigma_5, 1), (\sigma_5, 0)\} \notag\\
        C_1 &=\{(\sigma_1, 2), (\sigma_6, 0), (\sigma_7, 0), (\sigma_8,0)\} \notag\\
        C_2 &=\{(\sigma_9, 0)\} \label{eq:exCi} \\
        C_3 &=\{(\sigma_{10}, 0)\} \notag\\
        C_4 &=\emptyset. \notag
    \end{align}
    %
\end{exmp}


\section{Problem Formulation}\label{sec:problem}

Given the system model ${\bf G}$ in (\ref{eq:plant:model}), the $n$ security levels $\Sigma_0,\ldots,\Sigma_{n-1}$ in (\ref{eq:Sigmai}), and the $n+1$ cost levels $C_0,\ldots,C_n$ in (\ref{eq:C0})-(\ref{eq:Cn}), we formulate in this section two secret protection problems. 

To proceed, we need several definitions.
Let $u \geq 1$ be the least number of events that are required to be protected before any secret state may be reached from any system trajectory from the initial state. Also let $v \geq 0$ be the least security level that is needed for protecting the secrets. Write 
\begin{align} \label{eq:geq-v}
    \Sigma^{\geq v}_p := \bigdisjoint_{i=v}^{n-1} \Sigma_i
\end{align}
for the collection of protectable events where security levels are at least $v$.
The following definition formalizes the notion that the secret states are protected with at least $u$ number of protections with at least $v$ security level of protectable events.
\begin{defn}[$u-v-$secure reachability]\label{defn:u-reach:model}
    Consider a system $\mathbf{G}$ in \cref{eq:plant:model} with a set of secret states $Q_s$, the security level sets $\Sigma_i$ ($i\in [0, n-1]$) in (\ref{eq:Sigmai}), and let $u \geq
    1$, $v \geq 0$, and $\tilde{\Sigma}$ be a nonempty subset of $\Sigma_p^{\geq v}$ in (\ref{eq:geq-v}). We say that $Q_s$ is {\it reachable
    with at least $u$ protectable events of security level at least $v$ w.r.t. $\tilde{\Sigma}$} (or simply $Q_s$ is $u-v-$securely reachable) if the following condition holds:
    \begin{equation}\label{eq:condition:u-reach}
        (\forall s \in \Sigma^*) (\delta(q_0, s)! \sand \delta(q_0, s) \in
        Q_s) \implies s \in \underbrace{\Sigma^\ast \tilde{\Sigma}
        \Sigma^\ast \cdots \Sigma^\ast \tilde{\Sigma}
        \Sigma^\ast}_{\text{$\tilde{\Sigma}$ appears $u$ times}}.
    \end{equation}
\end{defn}
Condition~(\ref{eq:condition:u-reach}) means that every string from the initial state that can reach a secret state must contain at least $u$ protectable events of security level at least $v$.

Next we define a \emph{protection policy} that identifies which protectable events to protect at which states. Such a policy is what we aim to design for the system administrator.
\begin{defn}[protection policy]\label{defn:protectpolicy}
    For the system $\mathbf{G} = (Q, \Sigma = \Sigma_p \cup \Sigma_{up}, \delta, q_0, Q_m)$ in \cref{eq:plant:model}, a \emph{protection policy} $\mathcal{P}$ is a mapping that assigns to each state a subset of protectable events: 
    \begin{equation}\label{eq:protectpolicy}
        \mathcal{P}: Q \to \power(\Sigma_p)
    \end{equation}
    where $\power(\Sigma_p)$ denotes the power set of $\Sigma_p$.
\end{defn}
Note that what a protection policy specifies can also be interpreted as the
protection of a transition labeled by a protectable event at a given
state. For example, $\mathcal{P}(q) = \{\sigma_i, \sigma_j\}$ represents that
protectable events $\sigma_i$ and $\sigma_j$ occurring at state $q$ are
protected.

Now we are ready to formulate two secret protection problems studied in this paper. 
The first problem is to find a protection policy (if it exists) that protects all the secret states with at least a prescribed number of protections of at least a prescribed security level, and moreover the protection cost should be minimum.

\begin{prob}[Usability Aware Secret Securing with Multiple Protections and
Minumum Cost Problem, USCP]\label{prob:ssmcp}
    Consider a system $\mathbf{G}$ in \cref{eq:plant:model} with a set of secret states $Q_s$, the cost level sets $C_i$ ($i \in [0,n]$) in (\ref{eq:C0})-(\ref{eq:Cn}), and let $u \geq 1$, $v \geq 0$. 
    Find a protection policy $\mathcal{P}:
    Q \to \power(\Sigma_p)$ such that $Q_s$ is $u-v-$securely reachable and the index $i$ of $C_i$ is minimum.
\end{prob}

More generally, and this is typical in practice, secrets may have different importance. For example in online shopping systems, customers' credit card information is (likely) more important than their email address information (though the latter certainly also needs to be protected).
Thus the set of secret states $Q_s$ may be partitioned into $k \geq 1$ disjoint (nonempty) subsets $Q_{s1}, \cdots, Q_{sk}$; the level of importance rises as the index increases. 

Naturally the administrator wants to protect secrets of higher importance with events of higher security levels. Hence we associate each $Q_{sj}$ ($j \in [1,k]$) with a number $v_j$ that indicates the least security level required for protecting the secrets in $Q_{sj}$. These $v_j$ satisfy $0 \leq v_1 \leq \cdots \leq v_k (\leq n-1)$ according to the rising importance. With this additional consideration, we formulate our second problem.

\begin{prob}[Usability Aware Heterogeneous Secret Securing with Multiple Protections and
Minumum Cost Problem, UHSCP]\label{prob:hssmcp}
    Consider a system $\mathbf{G}$ in \cref{eq:plant:model}, a set of secret states $Q_s$ paritioned into disjoint (nonempty) subsets $Q_{s1}, \cdots, Q_{sk}$ with rising importance, the cost level sets $C_i$ ($i \in [0,n]$) in (\ref{eq:C0})-(\ref{eq:Cn}), and let $u \geq 1$, $0 \leq v_1 \leq \cdots \leq v_k \leq n-1$. 
    Find a protection policy $\mathcal{P}:
    Q \to \power(\Sigma_p)$ such that for every $j \in [1,k]$ the $j$th important secret state subset $Q_{sj}$ is $u-v_j-$securely reachable and the index $i$ of $C_i$ is minimum.
\end{prob}

  Let us revisit \cref{exmp:model} to explain the above formulated two problems.

\begin{exmp}
    Consider the system model $\mathbf{G}$ in \cref{fig:exmp:model:plant}, with the secret state set $Q_s = \{q_7, q_8, q_{10}\}$, the security level sets $\Sigma_i$ ($i \in [0, n-1]$) in (\ref{eq:exSigmai}),  
    and the cost level sets $C_i$ ($i\in [0, n]$) in (\ref{eq:exCi}). 
    
    For Problem~\ref{prob:ssmcp}, let $u=2$ and $v=0$; namely it is required that at least $2$ events be protected for every system trajectory (from the initial state) that may reach a secret state in $Q_s$, and the least security level is $0$. Then our goal is to find a protection policy $\mathcal{P}:
    Q \to \power(\Sigma_p)$ (if it exists) such that $Q_s$ is $2-0-$securely reachable, and moreover the index $i$ of $C_i$ is minimum (i.e. least cost).
     
    Next for Problem~\ref{prob:hssmcp}, 
    we consider that $Q_s$ is partitioned into two disjoint subsets $Q_{s1} = \{q_7,q_8\}$ and $Q_{s2} = \{q_{10}\}$. This means that $q_{10}$, the administrative realm, is a more important secret than $q_7$ and $q_8$ (regular users' secrets). Accordingly, let $v_1=0$ and $v_2=1$, namely the least security level for $Q_{s1}$ is $0$ while the least security level for $Q_{s2}$ is $1$; the latter means that when protecting the secret state $q_{10} \in Q_{s2}$, events $\sigma_0,\sigma_1,\sigma_5 \in \Sigma_0$ cannot be used due to their insufficient security level. 
    As for the required number of protections, we again let $u=2$. Then the objective here is to find a protection policy $\mathcal{P}:
    Q \to \power(\Sigma_p)$ (if it exists) such that $Q_{s1}$ is $2-0-$securely reachable, $Q_{s2}$ is $2-1-$securely reachable, and moreover the index $i$ of $C_i$ is minimum (i.e. least cost).
\end{exmp}


\section{Usability Aware Secret Securing with Minumum Cost}

In this section, we address  Problem~\ref{prob:ssmcp} (USCP). We start by characterizing the solvability of Problem~\ref{prob:ssmcp}, then present an algorithm to compute a solution, and finally illustrate the results using the running example (Example~2.1).

\subsection{Solvability of USCP}\label{subsec:solvability:uniform}

It is evident that if there are too few protectable events or the requirement for protection numbers and security levels is too high, then there might not exist a solution to Problem~\ref{prob:ssmcp}. 
The following theorem provides a necessary and sufficient condition under
which there exists a solution of Problem~\ref{prob:ssmcp}.

\begin{thm}\label{thm:solvable:uniform}
Consider a system $\mathbf{G}$ in \cref{eq:plant:model} with a set of secret states $Q_s$, the cost level sets $C_i$ ($i \in [0,n]$) in (\ref{eq:C0})-(\ref{eq:Cn}), the required least number of protections $u \geq 1$, and the required lowest security level $v \geq 0$. 
Problem~\ref{prob:ssmcp} is solvable (i.e.
there exists a protection policy $\mathcal{P}:
    Q \to \power(\Sigma_p)$ such that $Q_s$ is $u-v-$securely reachable and the index $i$ of $C_i$ is minimum)
if and only if
either 
    \begin{equation}\label{eq:thm:solvable:uniform:condition:1}
        \text{$Q_s$ is $u-0-$securely reachable w.r.t. $\tilde{\Sigma} = \Sigma(C_0)$};
    \end{equation}
    or there exists $i \in [v,n]$ such that
    \begin{equation}\label{eq:thm:solvable:uniform:condition:2}
        \begin{gathered}
        \text{$Q_s$ is $u-v-$securely reachable w.r.t. $\tilde{\Sigma} = \bigcup^i_{l = v} \Sigma(C_l) \setminus \Sigma_{v-1}$} \\
        \sand \\
        \text{$Q_s$ is not $u-v-$securely reachable w.r.t. $\tilde{\Sigma} = \bigcup^{i-1}_{l = v} \Sigma(C_l) \setminus \Sigma_{v-1}$}.
        \end{gathered}
    \end{equation}
\end{thm}

    Condition \cref{eq:thm:solvable:uniform:condition:1} means that in the special case where the required lowest security level $v =
    0$, every system trajectory reaching the secret states in $Q_s$ contains at least $u$ protectable events in $\Sigma(C_0) \subseteq \Sigma_0$. This is the easiest case, and the index $0$ is minimum. 
    
    More generally, condition
    \cref{eq:thm:solvable:uniform:condition:2} means that there exists an index $i \in [v, n]$ for which every system trajectory reaching the secret states in $Q_s$ contains at least $u$
    protectable events in $\bigcup^i_{l = v} \Sigma(C_l) \setminus \Sigma_{v-1} \subseteq \Sigma^{\geq v}_p$, but there exists at least one trajectory reaching $Q_s$ that contains fewer than $u$ 
    protectable events in $\bigcup^{i-1}_{l = v} \Sigma(C_l) \setminus \Sigma_{v-1} \subseteq \Sigma^{\geq v}_p$. That these two conditions in (\ref{eq:thm:solvable:uniform:condition:2}) simultaneously hold indicates that the index $i$ of the cost level sets $C_i$ is minimum. 
    Note that in \cref{eq:thm:solvable:uniform:condition:2} the set minus ``$\setminus \Sigma_{v-1}$'' is needed because $\Sigma(C_v) \subseteq \Sigma_{v-1} \cup \Sigma_{v}$ (as in (\ref{eq:SigmaC})), and the protectable events in $\Sigma_{v-1}$ do not satisfy the required security level $v$.

\begin{proof}
    ($\Rightarrow$) If condition~\cref{eq:thm:solvable:uniform:condition:1} holds,
    i.e. $Q_s$ is $u-0-$securely reachable w.r.t. $\Sigma(C_0) \subseteq \Sigma_0$, then the index $0$ is evidently the smallest. In this case, there exists a protection policy $\mathcal{P} : Q \to \power(\Sigma(C_0))$ as a
    solution for \cref{prob:ssmcp} using protectable events only in
    $\Sigma(C_0) \subseteq \Sigma_0$ which satisfies the required security level $0$. Therefore, if \cref{eq:thm:solvable:uniform:condition:1}
    holds, then \cref{prob:ssmcp} is solvable (for the special case $v=0$). 
    
    If
    \cref{eq:thm:solvable:uniform:condition:2} holds, then $Q_s$ is
    $u-v-$securely reachable w.r.t. $\bigcup^i_{l = v} \Sigma(C_l) \setminus \Sigma_{v-1}$, and moreover the index $i$ of $C_i$ is minimum. The latter is because $Q_s$ is not $u-v-$securely reachable w.r.t. $\bigcup^{i-1}_{l = v} \Sigma(C_l) \setminus \Sigma_{v-1}$ and 
    $\bigcup^{i-1}_{l = v} \Sigma(C_l) \setminus \Sigma_{v-1} \subseteq \bigcup^{i}_{l = v} \Sigma(C_l) \setminus \Sigma_{v-1}$.
    In this case, there exists a protection policy $\mathcal{P} : Q \to \power(\bigcup^{i}_{l = v} \Sigma(C_l) \setminus \Sigma_{v-1})$ as a solution for \cref{prob:ssmcp} using protectable
    events in $\bigcup^{i}_{l = v} \Sigma(C_l) \setminus \Sigma_{v-1} \subseteq \Sigma_p^{\geq v}$ which satisfies the required security level $v$. Therefore, if
    \cref{eq:thm:solvable:uniform:condition:2} holds, then \cref{prob:ssmcp}
    is solvable.

    ($\Leftarrow$) If \cref{prob:ssmcp} is solvable with the minimum index of $C_i$ being $i = 0$, then $Q_s$ is $u-0-$securely reachable w.r.t. $\Sigma(C_0)$. This is exactly condition~\cref{eq:thm:solvable:uniform:condition:1}. 
    
    If \cref{prob:ssmcp} is solvable with the minimum index of $C_i$ satisfying $v \leq i \leq n$, then
    $Q_s$ is $u-v-$securely reachable w.r.t. $\bigcup^i_{l = v} \Sigma(C_l) \setminus \Sigma_{v-1}$. Since the index $i$ is minimum, it indicates that
    $Q_s$ is not $u-v-$securely reachable w.r.t. $\bigcup^{i-1}_{l = v} \Sigma(C_l) \setminus \Sigma_{v-1}$. Therefore 
    \cref{eq:thm:solvable:uniform:condition:2} holds.
\end{proof}

\subsection{Policy Computation for USCP}\label{subsec:computation:uniform}

When \cref{prob:ssmcp} is solvable under the condition presented in Theorem~\ref{thm:solvable:uniform}, we design an algorithm to compute a solution, namely a protection policy.

To compute such a protection policy, our approach is to convert \cref{prob:ssmcp} (a security problem) to a corresponding control problem and adapt methods from the superviory control theory.

By this conversion, the sets of protectable events $\Sigma_p$ and
unprotectable events $\Sigma_{up}$ are interpreted as the sets of
\emph{controllable events} $\Sigma_c$ and \emph{uncontrollable events}
$\Sigma_{uc}$, respectively. Accordingly, a system $\mathbf{G}$ in
\cref{eq:plant:model} is changed to
\begin{equation}\label{eq:plant:control}
    \mathbf{G} = (Q, \Sigma, \delta, q_0, Q_m)
\end{equation}
where $\Sigma = \Sigma_c \disjoint \Sigma_{uc}$ and $\Sigma_c =
\bigdisjoint_{i=0}^{n-1} \Sigma_i$. Recall from (\ref{eq:Sigmai}) that $\Sigma_i$ ($i =
0, \ldots, n-1$) denote the partition of protectable events in $\Sigma_p$ as
the index $i$ represents the security level (and the first source of cost); accordingly, here $\Sigma_i$ denote the partition of controllable events in $\Sigma_c$. Similar to (\ref{eq:geq-v}), for a given $v \geq 0$ write
\begin{align} \label{eq:Sigmacv}
    \Sigma^{\geq v}_c := \bigdisjoint_{i=v}^{n-1} \Sigma_i.
\end{align}

In addition, protection policy
$\mathcal{P}: Q \to
\power(\Sigma_p)$ is changed to {\em control policy} $\mathcal{D}: Q \to
\power(\Sigma_c)$, which is a control decision (of a supervisor) specifying
which controllable events to disable at any given state. 
More specifically, let $\mathbf{S}
= (X, \Sigma, \xi, x_0, X_m)$ be a supervisor for system ${\bf G} = (Q,\Sigma,\delta,q_0,Q_m)$ and assume without loss of generality that ${\bf S}$ is a subautomaton of 
$\mathbf{G}$. The control policy $\mathcal{D}:Q \to
\power(\Sigma_c)$ is given by
\begin{equation}\label{eq:policy:control}
    \mathcal{D}(q) \coloneqq \begin{dcases}
        \{\sigma \in \Sigma_c \mid \neg \xi(q, \sigma)! \sand \delta(q, \sigma)!\}, & \text{if $q \in X$} \\
        \emptyset, & \text{if $q \in Q \setminus X$}
    \end{dcases}
\end{equation}

Based on the above conversion, \cref{defn:u-reach:model} and
\cref{prob:ssmcp} are changed to the following definition and problem.

\begin{defn}[$u-v-$controllable reachability]\label{defn:u-reach:control}
Consider a system $\mathbf{G}$ in \cref{eq:plant:control} with a set of secret states $Q_s$, the (security) level sets $\Sigma_i$ ($i\in [0, n-1]$) in (\ref{eq:Sigmai}), and let $u \geq
    1$, $v \geq 0$, and $\tilde{\Sigma}$ be a nonempty subset of $\Sigma_c^{\geq v}$ in (\ref{eq:Sigmacv}). We say that $Q_s$ is {\it reachable
    with at least $u$ controllable events of (security) level at least $v$ w.r.t. $\tilde{\Sigma}$} (or simply $Q_s$ is $u-v-$controllably reachable) if the following condition holds:
    \begin{equation}\label{eq:condition:u-reach:control}
        (\forall s \in \Sigma^*) (\delta(q_0, s)! \sand \delta(q_0, s) \in
        Q_s) \implies s \in \underbrace{\Sigma^\ast \tilde{\Sigma}
        \Sigma^\ast \cdots \Sigma^\ast \tilde{\Sigma}
        \Sigma^\ast}_{\text{$\tilde{\Sigma}$ appears $u$ times}}.
    \end{equation}
\end{defn}

\begin{prob}[Usability Aware Reachability Control with Multiple Controllable
Events and Minimum Cost Problem, UCCP]\label{prob:rcmcp}
    Consider a system $\mathbf{G}$ in \cref{eq:plant:control} with a set of secret states $Q_s$, the cost level sets $C_i$ ($i \in [0,n]$) in (\ref{eq:C0})-(\ref{eq:Cn}), and let $u \geq 1$, $v \geq 0$. 
    Find a control policy $\mathcal{P}:
    Q \to \power(\Sigma_c)$ such that $Q_s$ is $u-v-$controlably reachable and the index $i$ of $C_i$ is minimum.
\end{prob}

The solvability condition of \cref{prob:rcmcp}, stated in the corollary below, 
follows directly from \cref{thm:solvable:uniform} and the above presented conversion. 

\begin{cor}\label{prop:solvable:uniform}
Consider a system $\mathbf{G}$ in \cref{eq:plant:control} with a set of secret states $Q_s$, the cost level sets $C_i$ ($i \in [0,n]$) in (\ref{eq:C0})-(\ref{eq:Cn}), the required least number of protections $u \geq 1$, and the required lowest (security) level $v \geq 0$. 
Problem~\ref{prob:rcmcp} is solvable (i.e.
there exists a control policy $\mathcal{P}:
    Q \to \power(\Sigma_c)$ such that $Q_s$ is $u-v-$controllably reachable and the index $i$ of $C_i$ is minimum)
if and only if
either 
    \begin{equation}\label{eq:prop:solvable:uniform:condition:1}
        \text{$Q_s$ is $u-0-$controllably reachable w.r.t. $\tilde{\Sigma} = \Sigma(C_0)$};
    \end{equation}
    or there exists $i \in [v,n]$ such that
    \begin{equation}\label{eq:prop:solvable:uniform:condition:2}
        \begin{gathered}
        \text{$Q_s$ is $u-v-$controllably reachable w.r.t. $\tilde{\Sigma} = \bigcup^i_{l = v} \Sigma(C_l) \setminus \Sigma_{v-1}$} \\
        \sand \\
        \text{$Q_s$ is not $u-v-$controllably reachable w.r.t. $\tilde{\Sigma} = \bigcup^{i-1}_{l = v} \Sigma(C_l) \setminus \Sigma_{v-1}$}.
        \end{gathered}
    \end{equation}
\end{cor}

When Problem~\ref{prob:rcmcp} is solvable (equivalently Problem~\ref{prob:ssmcp} is solvable), we present an algorithm to compute a control policy as a solution for Problem~\ref{prob:rcmcp}. 
Such a control policy
specifies at least $u$ controllable events of (security) level at least $v$ to disable in every string from
the initial state $q_0$ to the secret state set $Q_s$. 
This control policy will finally be converted back to a protection policy as a solution for Problem~\ref{prob:ssmcp} (our original security problem).

The algorithm that we design to solve Problem~\ref{prob:rcmcp} is presented on the next page (Algorithm~1 UCC$u$). In the following we explain the main ingredients and steps of this algorithm. 

First, the inputs of Algorithm~1 are the system $\mathbf{G}$ in \cref{eq:plant:control}, a set of secret states $Q_s$, the least number of protections $u \geq 1$, and the least (security) level $v \geq 0$. Then Algorithm~1 will output $u$
supervisors $\mathbf{S}_0, \dots, \mathbf{S}_{u-1}$ for $\mathbf{G}$ (if they exist) as well as the minimum cost index $i_{\min}$. Each supervisor is computed by the UCC function (lines~14--24), and provides a different control policy such that every string
reaching secret states has at least one controllable event of (security) level at $v$.
So in total, $\mathbf{S}_0, \dots, \mathbf{S}_{u-1}$ specify $u$ controllable events to disable in every string reaching $Q_s$.

To compute the first supervisor $\mathbf{S}_0$, at line~1 of Algorithm~1 
we need to design the control specification $\mathbf{G}_K$. This is done by removing from ${\bf G}$ all the secret
states in $Q_s$ and the transition to and from the removed states. Hence
\begin{equation}\label{eq:spec:uniform}
    \mathbf{G}_K = (Q \setminus Q_s, \Sigma, \delta_K, q_0, Q \setminus Q_s)
\end{equation}
where $\delta_K = \delta \setminus \{(q, \sigma, q') \mid q \mbox{ or } q' \in Q_s,
\sigma \in \Sigma, \delta(q,\sigma)!, \delta(q,\sigma)=q'\}$.\footnote{Note that in real systems, secret states should still be reachable. Even though the computed supervisors specify
which controllable events to disable in the control context, we consider
the {\em protection} of these specified events so that secret states are still reachable
but protected. Our view is that in real systems, it is not desirable to disable controllable events and make secret states unreachable, because it would prevent regular users from ever accessing these secret states as well.} We remark that for ${\bf G}_K$ we let all of its states be marked; this is because we do not want to introduce extra control actions owing to ensuring nonblocking behavior. 

\begin{algorithm}[htp]
    \caption{UCC$u$}
    \label{alg:rcmc-u}
    \begin{algorithmic}[1]
        \Require{System $\mathbf{G}$, secret state set $Q_s$, protection number $u$, security level $v$}
        \Ensure{Supervisors $\mathbf{S}_0$, $\mathbf{S}_1$, \dots, $\mathbf{S}_{u-1}$, minimum cost index $i_{\min}$}
        \State{$\mathbf{G}_0 = (Q,\Sigma^0,\delta^0,q_0, Q_m) = \mathbf{G}, \mathbf{G}_{K,0} = \mathbf{G}_K$ as in (\ref{eq:spec:uniform})}%
        \For{$j = 0, 1, \dots, u-1$}%
        \State{$(\mathbf{S}_j, i_j) =$ \Call{UCC}{$\mathbf{G}_j$, $\mathbf{G}_{K,j}$, $v$}}%
        \If{$\mathbf{S}_j$ is nonempty}%
        \State{Derive $\mathcal{D}_j$ from $\mathbf{S}_j$ as in
        \cref{eq:policy:control}}%
        \State{Form $\mathbf{G}_{j+1} = (Q, \Sigma^{j+1}, \delta^{j+1}, q_0, Q_m)$ from $\mathbf{G}_j$ and $\mathcal{D}_j$ as in \cref{eq:plant:relabeled}}
        \State{$\delta_K^{j+1} = \delta^{j+1} \setminus \{(q, \sigma, q') \mid q \mbox{ or } q' \in Q_s,
\sigma \in \Sigma^{j+1},  \delta^{j+1}(q,\sigma)=q'\}$}
        \State{$\mathbf{G}_{K,j+1} = (Q \setminus Q_s, \Sigma^{j+1}, \delta_K^{j+1}, q_0,Q \setminus Q_s)$}
        \Else%
        \State\Return{Empty supervisors, index $-1$}%
        \EndIf%
        \EndFor%
        \State\Return{$\mathbf{S}_0$, $\mathbf{S}_1$, \dots, $\mathbf{S}_{u-1}$, $i_{\min} = i_{u-1}$}
        \Statex{}
        \Function{UCC}{$\mathbf{G}$, $\mathbf{G}_K$, $v$}%
        \State{$K = L(\mathbf{G}_K)$}%
        \For{$i = v, v+1, \dots, n$}%
        \State{$\Gamma = \bigcup^i_{l = v} \Sigma(C_l) \setminus \Sigma_{v-1}$}%
        \State{Compute a supervisor $\mathbf{S}$ s.t. $L(\mathbf{S}) = \supc(K)$
        w.r.t. ${\bf G}$ and $\Gamma$}%
        \If{$\mathbf{S}$ is nonempty}%
        \State\Return{$(\mathbf{S}, i)$}%
        \EndIf%
        \EndFor%
        \State\Return{(empty supervisor, index $-1$)}%
        \EndFunction%
    \end{algorithmic}
\end{algorithm}

\begin{exmp}\label{exmp:spec:uniform}
    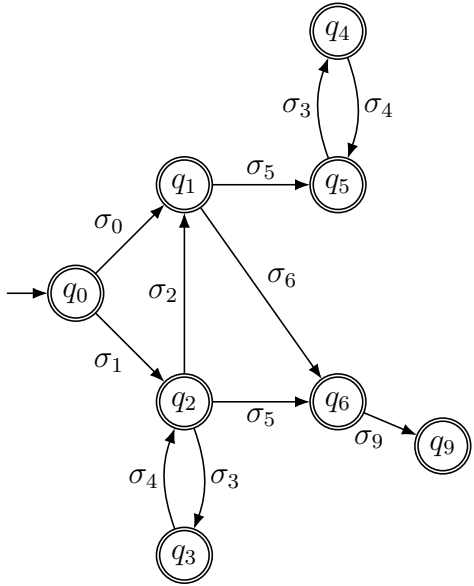
\begin{figure}[htp]
        \centering
        \adjustbox{}{
            \begin{tikzpicture}[automata]

\node[state, accepting, initial] (0) {$q_0$};
\node[state, accepting, above right=of 0] (1) {$q_1$};
\node[state, accepting, right=of 1] (2) {$q_5$};
\node[state, accepting, below right=of 0] (3) {$q_2$};
\node[state, accepting, right=of 3] (4) {$q_6$};
\node[state, accepting, below=of s2] (5) {$q_9$};
\node[state, accepting, below=of 3] (m1) {$q_3$};
\node[state, accepting, above=of 2] (m2) {$q_4$};

\path
(0) edge node{$\sigma_0$} (1)
(1) edge node{$\sigma_5$} (2)
(0) edge[swap] node{$\sigma_1$} (3)
(3) edge node{$\sigma_2$} (1)
(1) edge node{$\sigma_6$} (4)
(3) edge[swap] node{$\sigma_5$} (4)
(4) edge[swap] node{$\sigma_9$} (5)
(3) edge[bend left] node{$\sigma_3$} (m1)
(m1) edge[bend left] node{$\sigma_4$} (3)
(2) edge[bend left] node{$\sigma_3$} (m2)
(m2) edge[bend left] node{$\sigma_4$} (2)
;
\end{tikzpicture}
        }
        \caption{Specification automaton $\mathbf{G}_K$}
        \label{fig:exmp:spec:uniform}
    \end{figure}
    Displayed in \cref{fig:exmp:spec:uniform} is the 
    specification automaton ${\bf G}_K$  derived from the system ${\bf G}$ in \cref{exmp:model} and the secret state set $Q_s = \{q_7, q_8, q_{10}\}$. To design $\mathbf{G}_K$, secret states
    in $Q_s = \{q_7, q_8, q_{10}\}$ and transitions $(q_5, \sigma_7, q_7)$,
    $(q_5, \sigma_8, q_8)$, $(q_7, \sigma_8, q_8)$, $(q_8, \sigma_9, q_9)$
    and $(q_9, \sigma_{10}, q_{10})$ are removed from $\mathbf{G}$ in
    \cref{fig:exmp:model:plant} and all the states of $\mathbf{G}_K$ are marked.
\end{exmp}

With ${\bf G}_K$ constructed, line~2 of Algorithm~1 starts from $j=0$ and line~3 calls the RCMC function (with arguments ${\bf G}_0 = {\bf G}$, ${\bf G}_{K,0}={\bf G}_K$, $v$) to compute the first supervisor ${\bf S}_0$ and the minimum cost index $i_0$. 
To this end, several
standard concepts of supervisory control theory (SCT) \cite{wonham2019supervisory,Cai:2020,Wonham:2018} are employed and briefly reviewed below. 

Consider a system $\mathbf{G} =(Q,\Sigma=\Sigma_c \cup \Sigma_{uc}, \delta, q_0, Q_m)$ in
\cref{eq:plant:control}, and let $K = L(\mathbf{G}_K) \subseteq L(\mathbf{G})$
be a specification language derived from the specification automaton
$\mathbf{G}_K$ in (\ref{eq:spec:uniform}). 
For a subset of
the controllable events $\Gamma (\subseteq \Sigma_c)$, 
$K$ is said to be {\em controllable} with respect to $\mathbf{G}$ and
$\Gamma$ if $\prefix{K}(\Sigma \setminus \Gamma) \cap L(\mathbf{G}) \subseteq \prefix{K}$
where $\prefix{K}$ is the {\em prefix closure} of $K$.
We denote by the family $\mathcal{C}(K) \coloneqq \{K' \subseteq K \mid
\prefix{K}(\Sigma \setminus \Gamma) \cap L(\mathbf{G}) \subseteq
\prefix{K}\}$ the set of all controllable sublanguages of $K$ with respect to
$\mathbf{G}$ and $\Gamma$, and by $\supc(K) \coloneqq \bigcup\{K' \mid K' \in
\mathcal{C}(K)\}$ the supremal controllable sublanguage of $K$ with respect
to $\mathbf{G}$ and $\Gamma$ (which is known to always exist).

\begin{lem}\label{lem:supc}
    (cf.~\cite{wonham2019supervisory}) Consider a plant $\mathbf{G}=(Q,\Sigma=\Sigma_c \cup \Sigma_{uc}, \delta, q_0, Q_m)$ in
    \cref{eq:plant:control} and a specification language  $K \subseteq L(\mathbf{G})$. It holds that
    \begin{equation}\label{eq:lem:supc}
        \supc(K) = \emptyset~\text{(w.r.t. $\mathbf{G}$ and $\Sigma_{uc}$)}
        \iff (\exists s \in \Sigma_{uc}^*) s \in L(\mathbf{G}) \setminus K.
    \end{equation}
\end{lem}

From \cref{lem:supc} and the construction of $\mathbf{G}_K$ in
\cref{eq:spec:uniform}, letting $K = L(\mathbf{G}_K)$ and $i \in [v,n]$, we know that the first supervisor
$\mathbf{S}_0 = \supc(K)$ (with respect to $\mathbf{G}$ in
\cref{eq:plant:control} and $\bigcup^i_{l=v} \Sigma(C_l) \setminus \Sigma_{v-1}$) is nonempty if and only if every
string reaching the secret states in $Q_s$ from the initial state $q_0$ has
at least one controllable event belonging to $\bigcup^i_{l=v} \Sigma(C_l) \setminus \Sigma_{v-1}$. In other words,
$\supc(K) \neq \emptyset$ (with respect to $\mathbf{G}$ and $\bigcup^i_{l=v} \Sigma(C_l) \setminus \Sigma_{v-1}$)
if and only if 
\begin{align} \label{eq:nonemptysup}
\left( \forall s \in \left(\Sigma \setminus \left(\bigcup^i_{l=v} \Sigma(C_l) \setminus \Sigma_{v-1}\right) \right)^* \right) \delta(q_0,
s) \not\in Q_s.    
\end{align}
The computation of ${\bf S}_0$ is carried out in lines~15--22 of Algorithm~1. If a nonempty ${\bf S}_0$ is obtained (line~19; condition~(\ref{eq:nonemptysup}) holds), then it is returned together with the current index $i$ of the cost level sets (line~20). Since the index is incrementally increased (line~16), we know that the index $i$ in line~20 is minimum (for this is the first time that ${\bf S}_0$ is nonempty).

Once a nonempty supervisor ${\bf S}_j$ ($j \geq 0$) is obtained (line~4), Algorithm~1 proceeds to compute the next supervisor ${\bf S}_{j+1}$ (until we acquire $u$ nonempty supervisors).
To ensure that each supervisor 
provides a different control
policy (disabling different transitions) so as to meet the requirement of $u$ protections, we need to change the status of those transitions already disabled by ${\bf S}_j$ from controllable to uncontrollable, so that the next supervisor ${\bf S}_{j+1}$ is forced to disable other controllable transitions. 

This status change is done by event relabeling. Specifically, let     $\mathbf{G}_{j} = (Q, \Sigma^{j} = \Sigma_{uc,j} \dot\cup \Sigma_{c,j}, \delta^{j}, q_0, Q_m)$ be the $j$th system model and 
$\mathcal{D}_j$ be the control policy in \cref{eq:policy:control} corresponding to supervisor ${\bf S}_j$.
Then the set of controllable
transitions specified (or disabled) by $\mathcal{D}_j$ is 
\begin{align*}
    \delta_{\mathcal{D}_j} := \{(q, \sigma, q') \mid q \in Q \sand \sigma \in
    \mathcal{D}_j(q) \sand q' = \delta^j(q,\sigma)\}.
\end{align*}
We relabel the above transitions and obtain 
\begin{align*}
    \delta'_{\mathcal{D}_j} := \{(q, \sigma', q') \mid (q, \sigma, q') \in \delta_{\mathcal{D}_j} \sand \sigma' \notin \Sigma^j \}.
\end{align*}
Moreover, we designate these relabeled transition as uncontrollable, so the new uncontrollable event set is: 
\begin{align*}
\Sigma_{uc,j+1} = \Sigma_{uc,j} \disjoint{} \{\sigma' \mid (q,
    \sigma', q') \in \delta'_{\mathcal{D}_j}\}.
\end{align*}
On the other hand, the new controllable event set is: 
\begin{align*}
\Sigma_{c,j+1} = \Sigma_{c,j} \setminus \{\sigma \mid (\forall q \in Q) \delta^j(q,\sigma)! \sand \delta^j(q,\sigma) = q' \Rightarrow (q,
    \sigma, q') \in \delta_{\mathcal{D}_j}\}.
\end{align*}
In words, those controllable events whose corresponding transitions are all specified by $\mathcal{D}_j$ and therefore relabled no longer exist and are consequently removed from the controllable event set.
Therefore we obtain the new system model
\begin{equation}\label{eq:plant:relabeled}
    \mathbf{G}_{j+1} = (Q, \Sigma^{j+1}, \delta^{j+1}, q_0, Q_m)
\end{equation}
where
\begin{align}
    \Sigma^{j+1} &= \Sigma_{uc,j+1} \disjoint{} \Sigma_{c,j+1}\\
    \delta^{j+1} &= (\delta^j \setminus \delta_{\mathcal{D}_j}) \disjoint
    \delta'_{\mathcal{D}_j}.
\end{align}
The above is carried out in lines~5--6 of Algorithm~1. Moreover, lines~7--8 update the specification model ${\bf G}_{K,j+1}$ similar to (\ref{eq:spec:uniform}).

With the updated system ${\bf G}_{j+1}$ and specification ${\bf G}_{K,j+1}$, Algorithm~1 again calls the UCC function (line~3) to compute the next supervisor ${\bf S}_{j+1}$ and the corresponding minimum cost index $i_{j+1}$. This process continues until $j=u-1$, unless an empty supervisor is returned by the UCC function. In the latter case, Algorithm~1 returns empty supervisors and index $-1$. 

If \cref{alg:rcmc-u} succeeds to compute $u$ nonempty supervisors ${\bf S}_0, \ldots, {\bf S}_{u-1}$, then these supervisors will be returned, together with the minimum cost index $i_{\min} = \max(i_0,\ldots,i_{u-1})$ (line~13). It is evident from the above construction that the inequality chain $v \leq i_0 \leq \cdots \leq i_{u-1} \leq n$ holds; hence $i_{\min} = i_{u-1}$.

Let $\mathcal{D}_j$ be the control policy of $\mathbf{S}_j$ ($j=0,\ldots,u-1$). Then define the overall control policy $\mathcal{D}: Q \to \power(\Sigma_c)$ by taking the union of the controllable events specified by individual $\mathcal{D}_j$ at every state, namely
\begin{equation}\label{eq:merge:uniform}
    \mathcal{D}(q) = \bigcup_{j=0}^{u-1}\mathcal{D}_j(q),\quad q \in Q.
\end{equation}
Since each control policy $\mathcal{D}_j$ ($j \in [0, u-1]$) specifies controllable events such
that every string reaching secret states has at least one disabled event, $\mathcal{D}$ in \cref{eq:merge:uniform} specifies at least
$u$ controllable events to disable in every string reaching secret states
from the initial state. Moreover, it follows from line~16 of Algorithm~1 that the (security) level of all these $u$ events are at least $v$.

The time complexity of Algorithm~1 is $O(u(n-v)|Q|^2)$, where $u$ is from line~2, $n-v$ from line~16, and $|Q|^2$ from line~18. 
The correctness of Algorithm~1 is asserted in the following proposition.

\begin{prop}\label{prop:solution:urcmcp}
    \cref{alg:rcmc-u} (with inputs $\mathbf{G}$, $Q_s$, $u$ and $v$) returns
    $u$ nonempty supervisors and minimum cost index $i_{\min} (\in [v,n])$ if and only if \cref{prob:rcmcp} is solvable.
\end{prop}

\begin{proof}
By the aforementioned constructions in \cref{alg:rcmc-u}, in particular line~16 (incrementally increasing the index of cost level sets) and line~17 ($\bigcup^{i}_{l = v} \Sigma(C_l) \setminus \Sigma_{v-1}$ monotonically becoming larger as index $i$ increases), 
Algorithm~1 returns
    $u$ nonempty supervisors and minimum cost index $i_{\min} \in [v,n]$ if and only if either of the two conditions (\ref{eq:prop:solvable:uniform:condition:1}), (\ref{eq:prop:solvable:uniform:condition:2})
    holds. By Corollary~\ref{prop:solvable:uniform}, the latter is a necessary and sufficient condition for the solvability of \cref{prob:rcmcp}. Therefore our conclusion ensues.
\end{proof}

From the derived control policy $\mathcal{D}$ in \cref{eq:merge:uniform}, a solution for \cref{prob:ssmcp}, namely a protection policy
$\mathcal{P}: Q \to \power(\Sigma_p)$, is obtained by inverse conversion of
controllable events back to protectable events. In terms of $\mathcal{P}$, we
interpret disabled events by $\mathcal{D}$ as {\em protected events}. 

Finally, we state the main result in this section, which provides a solution to our original security protection problem USCP (\cref{prob:ssmcp}). 

\begin{thm}
    Consider a system $\mathbf{G}$ in \cref{eq:plant:model} with a set of secret states $Q_s$, the cost level sets $C_i$ ($i \in [0,n]$) in (\ref{eq:C0})-(\ref{eq:Cn}), the required least number of protections $u \geq 1$, and the required lowest security level $v \geq 0$. 
    If
    \cref{prob:ssmcp} is solvable, then the protection policy $\mathcal{P}$
    derived from $\mathcal{D}$ in \cref{eq:merge:uniform} is a solution.
\end{thm}

\begin{proof}
    Suppose that \cref{prob:ssmcp} is solvable. Then \cref{prob:rcmcp} is
    also solvable by conversion of protectable events to controllable events. Then
    by \cref{prop:solution:urcmcp}, \cref{alg:rcmc-u} returns $u$ nonempty supervisors and the minimum cost index $i_{\min} \in [v,n]$. Based on these $u$ supervisors, control policies $\mathcal{D}_0, \dots, \mathcal{D}_{u-1}$ may be derived as in \cref{eq:policy:control}. Hence, a combined control policy $\mathcal{D}$ in \cref{eq:merge:uniform} is obtained. Due to the event relabeling in (\ref{eq:plant:relabeled}), each control policy uniquely
    specifies transitions in $\mathbf{G}$ to disable. Also it follows from the specifications
    $\mathbf{G}_{K,0}, \dots, \mathbf{G}_{K,u-1}$ in \cref{alg:rcmc-u} that
    $Q_s$ is $1-v-$controllably reachable under each of $\mathcal{D}_0, \dots,
    \mathcal{D}_{u-1}$. Therefore, under control policy $\mathcal{D}$, $Q_s$
    is $u-v-$controllably reachable. Hence, the control policy $\mathcal{D}$ is a solution for
    \cref{prob:rcmcp}. Consequently, from the inverse conversion of controllable events back to protectable events, the protection
    policy $\mathcal{P}$ derived from $\mathcal{D}$ is a solution for
    \cref{prob:ssmcp}.
\end{proof}

\subsection{Running Example}\label{subsec:example:uniform}

Let us again use \cref{exmp:model} to demonstrate our developed solution via Algorithm~1 for
\cref{prob:ssmcp}. 

Consider the system $\mathbf{G}$ in \cref{fig:exmp:model:plant}, with the secret state set $Q_s = \{q_7, q_8, q_{10}\}$, the security level sets $\Sigma_i$ ($i \in [0, 3]$) in (\ref{eq:exSigmai}),  
    and the cost level sets $C_i$ ($i\in [0, 4]$) in (\ref{eq:exCi}). Let $u=2$ and $v=0$; namely it is required that at least $2$ events be protected for every system trajectory (from the initial state) that may reach a secret state in $Q_s$, and the least security level is $0$. We demonstrate how to use Algorithm~1 to compute a protection policy $\mathcal{P}:
    Q \to \power(\Sigma_p)$ and the minimum index $i$ of $C_i$ as a solution for \cref{prob:ssmcp}.

First, convert protectable events to controllable events such that 
    \begin{align*}
        \Sigma_c = \{\sigma_0,\sigma_1,\sigma_5,\sigma_6,\sigma_7,\sigma_8,\sigma_9,\sigma_{10}\}. 
    \end{align*}
Accordingly the uncontrollable event set $\Sigma_{uc} = \{\sigma_2,\sigma_3,\sigma_4\}$.
Then input Algorithm~1 with the converted
system model $\mathbf{G}$, $Q_s$, $u = 2$ and $v = 0$. 

In the first iteration ($j=0$), system $\mathbf{G}_0 = {\bf G}$ in \cref{fig:exmp:model:plant} and specification ${\bf G}_{K,0}=\mathbf{G}_K$ in
\cref{fig:exmp:spec:uniform}. Then the RCMC function is called to compute the first supervisor $\mathbf{S}_0$. It is verified that when $i=0$ (line~16), the supervisor ${\bf S}$ is empty (line~18), whereas when $i=1$, the supervisor ${\bf S}$ is nonempty. Thus this nonempty supervisor is returned as ${\bf S}_0$ and the index $1$ is returned as $i_0$ (line~20). 
The control policy
$\mathcal{D}_0$ correponding to $\mathbf{S}_0$ is:
\begin{align*}
    &\mathcal{D}_0(q_1) = \{\sigma_6\},\quad \mathcal{D}_0(q_2) = \{\sigma_5\},\quad \mathcal{D}_0(q_5) = \{\sigma_7, \sigma_8\},\\ 
    &(\forall q \in Q \setminus \{q_1,q_2,q_5\}) \mathcal{D}_0(q) = \emptyset.
\end{align*}
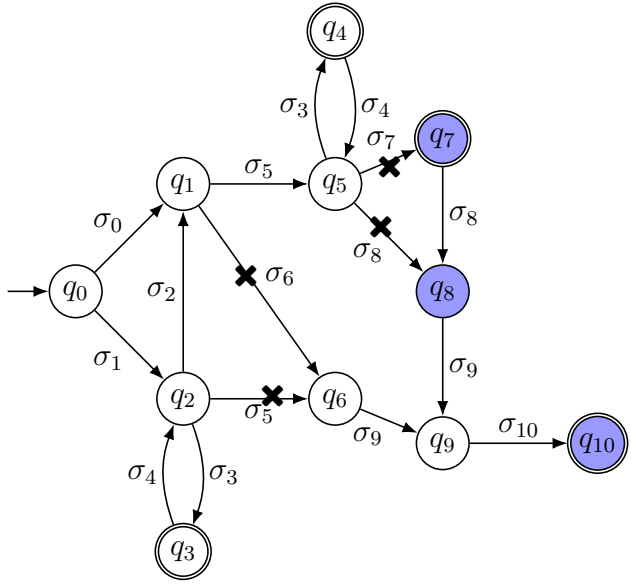
\begin{figure}[htp]
    \centering
    \adjustbox{}{%
        \begin{tikzpicture}[automata]

\node[state, initial] (0) {$q_0$};
\node[state, above right=of 0] (1) {$q_1$};
\node[state, right=of 1] (2) {$q_5$};
\node[state, below right=of 0] (3) {$q_2$};
\node[state, right=of 3] (4) {$q_6$};
\node[state, below right=of 2, secret1] (s2) {$q_8$};
\node[state, above=of s2, accepting, secret1] (s1) {$q_7$};
\node[state, below=of s2] (5) {$q_9$};
\node[state, right=of 5, accepting, secret1] (s3) {$q_{10}$};
\node[state, below=of 3, accepting] (m1) {$q_3$};
\node[state, above=of 2, accepting] (m2) {$q_4$};

\path
(0) edge node{$\sigma_0$} (1)
(1) edge node{$\sigma_5$} (2)
(0) edge[swap] node{$\sigma_1$} (3)
(3) edge node{$\sigma_2$} (1)
(1) edge node[label={[shift={(2pt,0pt)}]180:\faTimes}]{$\sigma_6$} (4)
(3) edge[swap] node[label={[shift={(-10pt,7pt)}]0:\faTimes}]{$\sigma_5$} (4)
(2) edge node[near end, label={[shift={(-12pt,-10pt)}]0:\faTimes}]{$\sigma_7$} (s1)
(2) edge[swap] node[label={[shift={(-10pt,10pt)}]0:\faTimes}]{$\sigma_8$} (s2)
(4) edge[swap] node{$\sigma_9$} (5)
(s1) edge node{$\sigma_8$} (s2)
(s2) edge node{$\sigma_9$} (5)
(5) edge node{$\sigma_{10}$} (s3)
(3) edge[bend left] node{$\sigma_3$} (m1)
(m1) edge[bend left] node{$\sigma_4$} (3)
(2) edge[bend left] node{$\sigma_3$} (m2)
(m2) edge[bend left] node{$\sigma_4$} (2)
;
\end{tikzpicture}
    }
    \caption{Control policy $\mathcal{D}_0$ of $\mathbf{S}_0$}\label{fig:poicy0:example:uniform}
\end{figure}
\cref{fig:poicy0:example:uniform} depicts the control policy $\mathcal{D}_0$
over the plant $\mathbf{G}$ in \cref{fig:exmp:model:plant}, indicating the
disabled transitions by ``\faTimes''.

We remark that since the lowest security level set is $\Sigma_0 = \{\sigma_0, \sigma_1, \sigma_5\}$, it would have been sufficient to disable $\sigma_0,\sigma_1$ at $q_0$ to satisfy the required $v=0$. However, disabling $\sigma_1$ would simultaneously affect regular users' accessing the (non-secret) marker states $q_3,q_4$, and this is deemed too costly in this example setting (threshold number is $T=2$ for the number of affected non-secret marker states). This observation makes it evident that taking into account the cost of usability generally requires the administrator to adopt a different protection policy. 

After obtaining $\mathcal{D}_0$, \cref{alg:rcmc-u} proceeds to relabel
the disabled transitions by $\mathcal{D}_0$ as follows:
\begin{align*}
    \delta_{\mathcal{D}_0} &= \{(q_1, \sigma_6, q_6), (q_2, \sigma_5, q_6), (q_5, \sigma_7, q_7), (q_5, \sigma_8, q_8)\} \\
    \delta_{\mathcal{D}_0}' &= \{(q_1, \sigma'_6, q_6), (q_2, \sigma'_5, q_6), (q_5, \sigma'_7, q_7), (q_5, \sigma'_8, q_8)\}.
\end{align*}
The relabeled events are designated to be uncontrollable events; thus the new uncontrollable event set is
\begin{align*}
    \Sigma_{uc,1} = \Sigma_{uc} \disjoint{}\{\sigma'_5, \sigma'_6, \sigma'_7, \sigma'_8\}.
\end{align*}
On the other hand, the new controllable event set is
\begin{align*}
    \Sigma_{c,1} = \Sigma_{c} \setminus \{\sigma_6, \sigma_7\}.
\end{align*}
Note that events $\sigma_5, \sigma_8$ remain in $\Sigma_{c,1}$ since they have other instances (of transitions) that are not disabled by $\mathcal{D}_0$.
From the above, the new system becomes $\mathbf{G}_1 = (Q, \Sigma^1, \delta^1, q_0, Q_m)$ where
\begin{align*}
    \Sigma^1 = \Sigma_{uc,1} \disjoint \Sigma_{c,1},\quad
    \delta^1 = (\delta \setminus \delta_{\mathcal{D}_0}) \disjoint \delta_{\mathcal{D}_0}'
\end{align*}
and the new specification automaton becomes
\begin{align*}
    \mathbf{G}_{K,1} = (Q \setminus Q_s, \Sigma^1, \delta^1_K, q_0, Q \setminus Q_s)
\end{align*}
where
\begin{align*}
    \delta^1_K = \delta^1 \setminus \{(q, \sigma, q') \mid q \mbox{ or } q' \in Q_s, \sigma \in \Sigma^1, \delta^1(q,\sigma)=q'\}.
\end{align*}
The new system $\mathbf{G}_1$ and specification $\mathbf{G}_{K,1}$ are displayed in
\cref{fig:relabeled:example:uniform} and \cref{fig:spec1:example:uniform},
respectively.
\begin{figure}[htp]
    \centering
    \adjustbox{}{%
        \begin{tikzpicture}[automata]
\node[state, initial] (0) {$q_0$};
\node[state, above right=of 0] (1) {$q_1$};
\node[state, right=of 1] (2) {$q_5$};
\node[state, below right=of 0] (3) {$q_2$};
\node[state, right=of 3] (4) {$q_6$};
\node[state, below right=of 2, secret1] (s2) {$q_8$};
\node[state, above=of s2, accepting, secret1] (s1) {$q_7$};
\node[state, below=of s2] (5) {$q_9$};
\node[state, right=of 5, accepting, secret1] (s3) {$q_{10}$};
\node[state, below=of 3, accepting] (m1) {$q_3$};
\node[state, above=of 2, accepting] (m2) {$q_4$};

\path
(0) edge node{$\sigma_0$} (1)
(1) edge node{$\sigma_5$} (2)
(0) edge[swap] node{$\sigma_1$} (3)
(3) edge node{$\sigma_2$} (1)
(1) edge node{$\sigma'_6$} (4)
(3) edge[swap] node{$\sigma'_5$} (4)
(2) edge node[very near end]{$\sigma'_7$} (s1)
(2) edge[swap] node{$\sigma'_8$} (s2)
(4) edge[swap] node{$\sigma_9$} (5)
(s1) edge node{$\sigma_8$} (s2)
(s2) edge node{$\sigma_9$} (5)
(5) edge node{$\sigma_{10}$} (s3)
(3) edge[bend left] node{$\sigma_3$} (m1)
(m1) edge[bend left] node{$\sigma_4$} (3)
(2) edge[bend left] node{$\sigma_3$} (m2)
(m2) edge[bend left] node{$\sigma_4$} (2)
;
\end{tikzpicture}
    }
    \caption{Relabeled system $\mathbf{G}_1$}\label{fig:relabeled:example:uniform}
\end{figure}
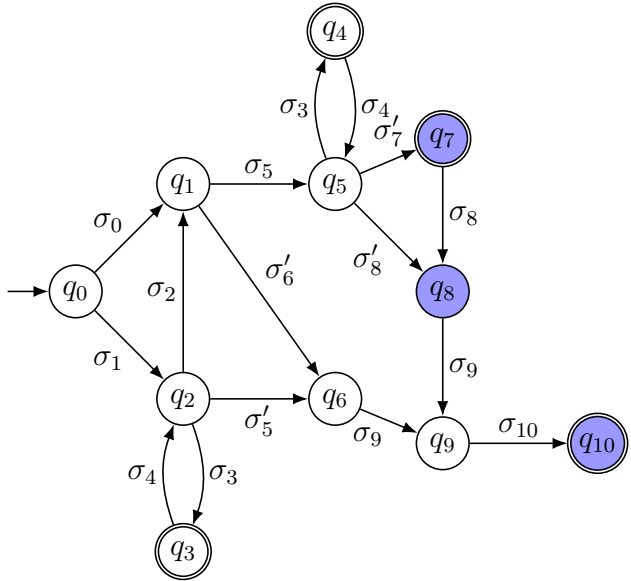
\begin{figure}[htp]
    \centering
    \adjustbox{}{%
        \begin{tikzpicture}[automata]
\node[state, accepting, initial] (0) {$q_0$};
\node[state, accepting, above right=of 0] (1) {$q_1$};
\node[state, accepting, right=of 1] (2) {$q_5$};
\node[state, accepting, below right=of 0] (3) {$q_2$};
\node[state, accepting, right=of 3] (4) {$q_6$};
\node[state, accepting, below=of s2] (5) {$q_9$};
\node[state, accepting, below=of 3] (m1) {$q_3$};
\node[state, accepting, above=of 2] (m2) {$q_4$};

\path
(0) edge node{$\sigma_0$} (1)
(1) edge node{$\sigma_5$} (2)
(0) edge[swap] node{$\sigma_1$} (3)
(3) edge node{$\sigma_2$} (1)
(1) edge node{$\sigma'_6$} (4)
(3) edge[swap] node{$\sigma'_5$} (4)
(4) edge[swap] node{$\sigma_9$} (5)
(3) edge[bend left] node{$\sigma_3$} (m1)
(m1) edge[bend left] node{$\sigma_4$} (3)
(2) edge[bend left] node{$\sigma_3$} (m2)
(m2) edge[bend left] node{$\sigma_4$} (2)
;
\end{tikzpicture}
    }
    \caption{Updated specification $\mathbf{G}_{K,1}$}
    \label{fig:spec1:example:uniform}
\end{figure}
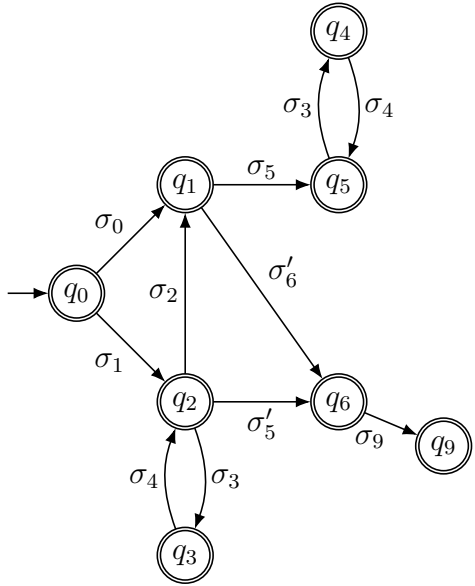

With $\mathbf{G}_1$ and $\mathbf{G}_{K,1}$, \cref{alg:rcmc-u} in the second iteration ($j=1$) again calls the RCMC function to compute the second supervisor $\mathbf{S}_1$. Like in the first iteration, when $i=0$ (line~16) the supervisor ${\bf S}$ is empty (line~18), whereas when $i=1$ the supervisor ${\bf S}$ is nonempty. Thus this nonempty supervisor is returned as ${\bf S}_1$ and the index $1$ is returned as $i_1$ (line~20). 
The control policy
$\mathcal{D}_1$ correponding to $\mathbf{S}_1$ is:
\begin{align*}
    \mathcal{D}_0(q_0) = \{\sigma_0, \sigma_1\},\quad
    (\forall q \in Q \setminus \{q_0\}) \mathcal{D}_0(q) = \emptyset.
\end{align*}

By now Algorithm~1 has succeeded in computing two nonempty supervisors. Since $u=2$, Algorithm~1 terminates and returns ${\bf S}_0$, ${\bf S}_1$, and the minimum cost index $i_{\min}=i_1=1$. 
Now we combine the two corresponding control policies into $\mathcal{D}$ as follows:
\begin{align*}
    \mathcal{D}(q) = \begin{dcases}
        \{\sigma_0, \sigma_1\}, & \text{if $q = q_0$} \\
        \{\sigma_6\}, & \text{if $q = q_1$} \\
                \{\sigma_5\}, & \text{if $q = q_2$} \\
        \{\sigma_7,\sigma_8\}, & \text{if $q = q_5$} \\
        \emptyset, & \text{if $q \in Q \setminus \{q_0, q_1,q_2, q_5\}$}
    \end{dcases}
\end{align*}
This $\mathcal{D}$ is a solution of \cref{prob:rcmcp}.

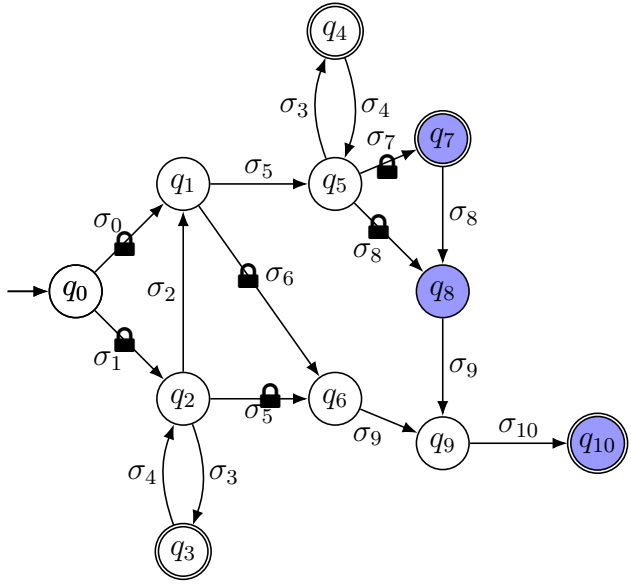
\begin{figure}[htp]
    \centering
    \adjustbox{}{%
        \begin{tikzpicture}[automata]
\node[state, initial] (0) {$q_0$};
\node[state, initial] (0) {$q_0$};
\node[state, above right=of 0] (1) {$q_1$};
\node[state, right=of 1] (2) {$q_5$};
\node[state, below right=of 0] (3) {$q_2$};
\node[state, right=of 3] (4) {$q_6$};
\node[state, below right=of 2, secret1] (s2) {$q_8$};
\node[state, above=of s2, accepting, secret1] (s1) {$q_7$};
\node[state, below=of s2] (5) {$q_9$};
\node[state, right=of 5, accepting, secret1] (s3) {$q_{10}$};
\node[state, below=of 3, accepting] (m1) {$q_3$};
\node[state, above=of 2, accepting] (m2) {$q_4$};

\path
(0) edge node[label={[shift={(-8pt,-8pt)}]0:\faLock}]{$\sigma_0$} (1)
(1) edge node{$\sigma_5$} (2)
(0) edge[swap] node[label={[shift={(-8pt,8pt)}]0:\faLock}]{$\sigma_1$} (3)
(3) edge node{$\sigma_2$} (1)
(1) edge node[label={[shift={(2pt,0pt)}]180:\faLock}]{$\sigma_6$} (4)
(3) edge[swap] node[label={[shift={(-10pt,7pt)}]0:\faLock}]{$\sigma_5$} (4)
(2) edge node[near end, label={[shift={(-12pt,-10pt)}]0:\faLock}]{$\sigma_7$} (s1)
(2) edge[swap] node[label={[shift={(-10pt,10pt)}]0:\faLock}]{$\sigma_8$} (s2)
(4) edge[swap] node{$\sigma_9$} (5)
(s1) edge node{$\sigma_8$} (s2)
(s2) edge node{$\sigma_9$} (5)
(5) edge node{$\sigma_{10}$} (s3)
(3) edge[bend left] node{$\sigma_3$} (m1)
(m1) edge[bend left] node{$\sigma_4$} (3)
(2) edge[bend left] node{$\sigma_3$} (m2)
(m2) edge[bend left] node{$\sigma_4$} (2)
;
\end{tikzpicture}
    }
    \caption{Protection policy $\mathcal{P}$ for $\mathbf{G}$}
    \label{fig:solution:example:uniform}
\end{figure}

Finally, by inverse conversion of controllable events back to protectable evvents we obtain a corresponding protection policy
$\mathcal{P}$ as a solution of the original \cref{prob:ssmcp}. 
\cref{fig:solution:example:uniform} illustrates this protection policy $\mathcal{P}$, where
``\faLock'' means the transitions that need to be ``protected''. 

Observe that based on this protetion policy $\mathcal{P}$,
every string from $q_0$ that can reach the secret states in $Q_s$ has at least two protected events in $\Sigma(C_0) \cup \Sigma(C_1) \subseteq \Sigma_0 \cup \Sigma_1$. Thus the least number of protections $u=2$ and the lowest security level $v=0$ are satisfied; moreover, the minimum cost index is $i_{\min} =1$.

For this example, the protections of each protected event specified by the policy $\mathcal{P}$ may be implemented as follows:
\begin{itemize}
    \item $\sigma_0$, $\sigma_1$: setting up a password on each account of the
    regular user and the administrator.
    \item $\sigma_5$: setting up a password for launching the application. 
    \item $\sigma_6$: setting up one-time password authentification.
    \item $\sigma_7$, $\sigma_8$: setting up fingerprint authentication.
\end{itemize}

\section{Usability Aware Heterogeneous Secret Securing with Minimum Cost}\label{sec:multilevel}

In this section, we move on to address Problem~\ref{prob:hssmcp} (UHSCP), in which the set of secret states
$Q_s$ is partitioned into $k (\geq 1)$   groups $Q_{s1},\ldots,Q_{sk}$ with heterogeneous importance; as the index $j \in [1,k]$ increases, the importance of $Q_{sj}$ rises. Similar to the preceding section, we begin with a characterization of the solvability of Problem~\ref{prob:hssmcp}, then present a solution algorithm, and finally use our running example to illustrate the results.

\subsection{Solvability of UHSCP}\label{subsec:solvability:multilevel}

The following theorem provides a necessary and sufficient condition under
which there exists a solution to Problem~\ref{prob:hssmcp}.

\begin{thm}\label{thm:solvable:multilevel}
Consider a system $\mathbf{G}$ in \cref{eq:plant:model}, a set of secret states $Q_s = \dot{\bigcup}^k_{j = 1} Q_{sj}$, the cost level sets $C_i$ ($i \in [0,n]$) in (\ref{eq:C0})-(\ref{eq:Cn}), the required least number of protections $u \geq 1$, and the required lowest security levels $v_j \geq 0$ for $Q_{sj}$ such that $v_1 \leq \cdots \leq v_k$. 
Problem~\ref{prob:hssmcp} is solvable (i.e.
there exists a protection policy $\mathcal{P}:
    Q \to \power(\Sigma_p)$ such that for every $j \in [1,k]$, $Q_{sj}$ is $u-v_j-$securely reachable and the index $i$ of $C_i$ is minimum)
if and only if
there exists $i \in [v_1,n]$ such that
    \begin{equation} \label{eq:thm:solvable:multilevel:condition}
        \begin{gathered}
        \text{$(\forall j \in [1,k]) Q_{sj}$ is $u-v_j-$securely reachable w.r.t. $\tilde{\Sigma}_j = \bigcup^i_{l = v_j} \Sigma(C_l) \setminus \Sigma_{v_j-1}$} \\
        \sand \\
        \text{$(\exists j \in [1,k]) Q_{sj}$ is not $u-v_j-$securely reachable w.r.t. $\tilde{\Sigma}_j = \bigcup^{i-1}_{l = v_j} \Sigma(C_l) \setminus \Sigma_{v_j-1}$}.
        \end{gathered}
    \end{equation}
\end{thm}

    Condition
    \cref{eq:thm:solvable:multilevel:condition} means that there exists an index $i \in [v_1, n]$ such that for every $j \in [1,k]$, the secret states in $Q_{sj}$ can be protected with at least $u$
    protections using protectable events in $\bigcup^i_{l = v_j} \Sigma(C_l) \setminus \Sigma_{v_j-1} \subseteq \Sigma^{\geq v_j}_p$, but there is $j \in [1,k]$ such that if only
    protectable events in $\bigcup^{i-1}_{l = v_j} \Sigma(C_l) \setminus \Sigma_{v_j-1} \subseteq \Sigma^{\geq v_j}_p$ are used, secrets cannot be
    protected with $u$ protections. That these two conditions in (\ref{eq:thm:solvable:multilevel:condition}) simultaneously hold indicates that the cost level index $i$ is minimum. 

\begin{proof}
     ($\Rightarrow$) If condition~\cref{eq:thm:solvable:multilevel:condition} holds, then for every $j \in [1,k]$, the secret subset $Q_{sj}$ is
    $u-v_j-$securely reachable w.r.t. $\bigcup^i_{l = v_j} \Sigma(C_l) \setminus \Sigma_{v_j-1}$, and moreover the index $i$ of $C_i$ is minimum. The latter is because at least one secret subset $Q_{sj'}$ ($j' \in [1,k]$) is not $u-v_{j'}-$securely reachable w.r.t. $\bigcup^{i-1}_{l = v_{j'}} \Sigma(C_l) \setminus \Sigma_{v_{j'}-1}$ and 
    $\bigcup^{i-1}_{l = v_{j'}} \Sigma(C_l) \setminus \Sigma_{v_{j'}-1} \subseteq \bigcup^{i}_{l = v_{j'}} \Sigma(C_l) \setminus \Sigma_{v_{j'}-1}$. 
    In this case, for every $Q_{sj}$ there exists a protection policy $\mathcal{P}_j : Q \to \power(\bigcup^{i}_{l = v_j} \Sigma(C_l) \setminus \Sigma_{v_j-1})$ such that protectable
    events in $\bigcup^{i}_{l = v_j} \Sigma(C_l) \setminus \Sigma_{v_j-1}$ may be used to satisfy the required least number of protections $u$ and the lowest security level $v_j$. These protection policies $\mathcal{P}_j$ ($j \in [1,k]$) together comprise a solution for \cref{prob:ssmcp}. Therefore, if
    \cref{eq:thm:solvable:multilevel:condition} holds, then \cref{prob:hssmcp}
    is solvable.

    ($\Leftarrow$) If \cref{prob:hssmcp} is solvable with the minimum index of $C_i$ being $i \in [v_1, n]$, then
    for every $j\in [1,k]$, $Q_{sj}$ is $u-v_j-$securely reachable w.r.t. $\bigcup^i_{l = v_j} \Sigma(C_l) \setminus \Sigma_{v_j-1}$. Since the index $i$ is minimum, it indicates that there exists at least one $j' \in [1,k]$ such that
    $Q_{sj'}$ is not $u-v_{j'}-$securely reachable w.r.t. $\bigcup^{i-1}_{l = v_{j'}} \Sigma(C_l) \setminus \Sigma_{v_{j'}-1}$. Therefore 
    \cref{eq:thm:solvable:multilevel:condition} holds.
\end{proof}

\subsection{Policy Computation for UHSCP}\label{subsec:computation:multilevel}

When \cref{prob:hssmcp} is solvable under the condition presented in Theorem~\ref{thm:solvable:multilevel}, we design an algorithm to compute a solution protection policy.

To compute such a protection policy, like in Section~4.2 we again convert the security problem to a corresponding control problem by changing protectable events to controllable events. Then we employ Algorithm~1 to compute a control policy for each secret subset $Q_{sj}$ ($j \in [1,k]$) to satisfy the required least number of protections $u$ and the lowest security level $v_j$. This is done by inputting Algorithm~1 with ${\bf G}$ in (\ref{eq:plant:control}), $Q_{sj}$, $u$ and $v_j$. 

If a solution exists, Algorithm~1 outputs $u$ supervisors ${\bf S}_{0,j},\ldots,{\bf S}_{u-1,j}$ and the minimum cost index $i_{\min, j}$. 
For these supervisors, one obtains 
the corresponding control policies $\mathcal{D}_{0,j},\ldots,\mathcal{D}_{u-1,j}$, which may be combined into a single control policy
\begin{align} \label{eq:Dj}
    \mathcal{D}_j(q) = \bigcup_{l=1}^{u-1} \mathcal{D}_{l,j}(q),\quad q \in Q.
\end{align}
If the above holds for all $j \in [1,k]$, further combining all resulting $\mathcal{D}_j$ ($j \in [1,k]$) yields an overall control policy $\mathcal{D}$ as follows:
\begin{equation}\label{eq:merge:multilevel}
    \mathcal{D}(q) = \bigcup_{j=1}^{k}\mathcal{D}_j(q), \quad q \in Q.
\end{equation}
One the other hand, the overall minimum cost index $i_{\min}$ satisfies:
\begin{align*}
    i_{\min} = \max(i_{\min, 1},\ldots,i_{\min, k} ).
\end{align*}


\begin{algorithm}[htp]
    \caption{UHCC$u$}\label{alg:2-mrcmc}
    \begin{algorithmic}[1]
        \Require{System ${\bf G}$ in (\ref{eq:plant:control}), secret state set $Q_s = \dot{\bigcup}_{j=1}^k Q_{sj}$, protection number $u$, security levels $0 \leq v_1 \leq \cdots \leq v_k \leq n$. }
        \Ensure{Control policy $\mathcal{D}$, minimum cost index $i_{\min}$}
        \For{$j = 1,\ldots,k$}
        \State{$\mathbf{S}_{0,j}, \dots, \mathbf{S}_{u-1,j}, i_{\min,j} =$
        \Call{UCC$u$}{$\mathbf{G}$, $Q_{sj}$, $u$, $v_{j}$}}
        \If{all $\mathbf{S}_{0,j}, \dots, \mathbf{S}_{u-1,j}$ are nonempty (or equivalently $i_{\min,j} \neq -1$)}
        \State{Derive $\mathcal{D}_j$ from
        $\mathbf{S}_{0,j}, \dots, \mathbf{S}_{u-1,j}$ as in
        (\ref{eq:Dj})}%
        \EndIf%
        \EndFor%
        \If{all $i_{\min,1},\ldots,i_{\min,k}$ are not equal to $-1$}
        \State{Derive $\mathcal{D}$ from $\mathcal{D}_{1},\ldots,\mathcal{D}_k$ as in (\ref{eq:merge:multilevel})}
        \State\Return{$\mathcal{D}$ and $i_{\min} = \max(i_{\min, 1},\ldots,i_{\min, k} )$}%
        \EndIf%
        \State\Return{Empty control policy $\mathcal{D}$ and index $-1$}
    \end{algorithmic}
\end{algorithm}

The above procedure is summarized in Algorithm~2 UHCC$u$. 
The time complexity of Algorithm~2 is $k$ (from line~1 and $k$ is the number of heterogeneoous secret subsets) times that of Algorithm~1, namely  $O(ku(n-v_1)|Q|^2)$. 
In fact, the $k$ calls to Algorithm~1 in line~2 can be done independently; hence the $k$ executions of lines~2--5 may be implemented on multi-core processors in a distributed (thus more efficient) manner.

If Algorithm~2 successfully outputs a (nonempty) control policy $\mathcal{D}$, then we convert it to a protection policy $\mathcal{P}: Q \to \power(\Sigma_p)$ by changing all controllable events back to protectable events. In terms of $\mathcal{P}$, we
interpret disabled events by $\mathcal{D}$ as {\em protected events}. 

Our main result in this section below asserts that the converted protection policy $\mathcal{P}$ is a solution for our original security problem UHSCP (\cref{prob:hssmcp}). 

\begin{thm}
    Consider a system $\mathbf{G}$ in \cref{eq:plant:model}, a set of secret states $Q_s = \dot{\bigcup}^k_{j = 1} Q_{sj}$, the cost level sets $C_i$ ($i \in [0,n]$) in (\ref{eq:C0})-(\ref{eq:Cn}), the required least number of protections $u \geq 1$, and the required lowest security levels $v_j \geq 0$ for $Q_{sj}$ such that $v_1 \leq \cdots \leq v_k$. 
    If
    \cref{prob:hssmcp} is solvable, then the protection policy $\mathcal{P}$
    derived from $\mathcal{D}$ in (\ref{eq:merge:multilevel}) (computed by Algorithm~2) is a solution.
\end{thm}

\begin{proof}
    Suppose that \cref{prob:hssmcp} is solvable. Then it follows from Theorem~\ref{thm:solvable:multilevel} that (\ref{eq:thm:solvable:multilevel:condition}) holds, i.e. there is $i \in [v_1,n]$ such that the two conditions in (\ref{eq:thm:solvable:multilevel:condition}) are satisfied. 
    
    Convert all protectable events to controllable events. The first condition in (\ref{eq:thm:solvable:multilevel:condition}) ensures that Algorithm~2 passes the test in line~3 for all $j \in [1,k]$. Hence, $k$ control policies $\mathcal{D}_j$ ($j \in [1,k]$) are obtained, each $\mathcal{D}_j$ ensuring that the secret subset $Q_{sj}$ is protected by $u$ protections, and the lowest security level of these protections is $v_j$. Again by the first condition in (\ref{eq:thm:solvable:multilevel:condition}), Algorithm~2 passes the test in line~7 and a combined control policy $\mathcal{D}$ is obtained from $\mathcal{D}_j$ ($j \in [1,k]$). Converting all controllable events back to protectable events, we derive the corresponding protection policy $\mathcal{P}$ which ensures $u-v_j-$secure reachability of $Q_{sj}$ for all $j \in [1,k]$.
    
    Finally, since each index $i_{\min,j}$ ($j \in [1,k]$) is minimum for the respective call to UCC$u$(${\bf G}, Q_{sj}, u, v_j$) and $i_{\min} = \max_{j \in [1,k]} i_{\min,j}$, it follows from the second condition in (\ref{eq:thm:solvable:multilevel:condition}) that $i_{\min}$ is the minimum cost index for the derived protection policy $\mathcal{P}$ as a solution for \cref{prob:hssmcp}.
\end{proof}

\subsection{Running Example}\label{subsec:example:multilevel}

For illustration let us revisit \cref{exmp:model}.
Consider the system $\mathbf{G}$ in
\cref{fig:exmp:model:plant}, with the secret state set $Q_s$ partitioned into two subsets: $Q_{s1} = \{q_7, q_8\}$ (regular users' secrets) and $Q_{s2}=\{q_{10}\}$ (administrator's secret). 
Accordingly, we require the lowest security levels to be $v_1 = 0$ and $v_2=1$, respectively. 
For the required number of protections, we let $u=2$ (the same as Section~4.3).

In addition, the security level sets are $\Sigma_i$ ($i \in [0, 3]$) as in (\ref{eq:exSigmai}), and the cost level sets are $C_i$ ($i\in [0, 4]$) as in (\ref{eq:exCi}). We demonstrate how to use Algorithm~2 to compute a protection policy $\mathcal{P}:
    Q \to \power(\Sigma_p)$ and the minimum index $i$ of $C_i$ as a solution for \cref{prob:hssmcp}.

First, convert protectable events to controllable events and input Algorithm~2 with the converted ${\bf G}$, $Q_s = Q_{s1} \dot{\cup} Q_{s2}$, $u=2$, $v_1=0$ and $v_2=1$. 

For $j=1$, call UCC$u$(${\bf G}, Q_{s1}, u, v_1$) to compute $u$ (nonempty) supervisors $\mathbf{S}_{0,1}, \dots, \mathbf{S}_{u-1,1}$ and the minimum cost index $i_{\min,1}=1$. From these supervisors, we obtain the corresponding control policy $\mathcal{D}_1$ as in (\ref{eq:Dj}):
\begin{align*}
    \mathcal{D}_1(q) = \begin{dcases}
        \{\sigma_5\}, & \text{if $q = q_1$} \\
        \{\sigma_7, \sigma_8\}, & \text{if $q = q_5$} \\
        \emptyset, & \text{if $q \in Q \setminus \{q_1,q_5\}$}
    \end{dcases}
\end{align*} 

Similarly for $j=2$, call UCC$u$(${\bf G}, Q_{s2}, u, v_2$) to compute $u$ (nonempty) supervisors $\mathbf{S}_{0,2}, \dots, \mathbf{S}_{u-1,2}$ and the minimum cost index $i_{\min,2}=3$. From these supervisors, we obtain the corresponding control policy $\mathcal{D}_2$ as in (\ref{eq:Dj}):
\begin{align*}
    \mathcal{D}_2(q) = \begin{dcases}
        \{\sigma_9\}, & \text{if $q = q_6$} \\
        \{\sigma_9\}, & \text{if $q = q_8$} \\
        \{\sigma_{10}\}, & \text{if $q = q_9$} \\
        \emptyset, & \text{if $q \in Q \setminus \{q_6, q_8, q_9\}$}
    \end{dcases}
\end{align*} 
It is interesting to observe that due to the required lowest security level $v_2=1$, events in $\Sigma_0=\{\sigma_0,\sigma_1,\sigma_5\}$ cannot be used (even though the event $\sigma_1$ at state $q_0$ belongs to $\Sigma(C_1)$). Consequently in this example, the events in the highest two security levels $\Sigma_2, \Sigma_3$ have to be used in order to meet this requirement.  

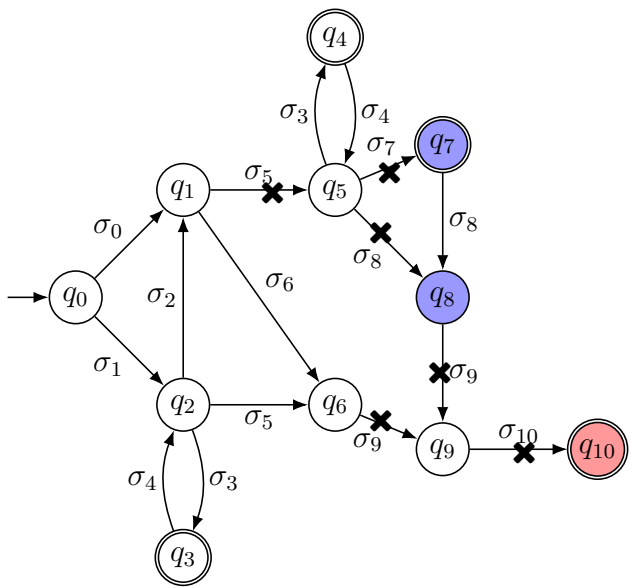
\begin{figure}[htp]
    \centering
    \adjustbox{}{%
        \begin{tikzpicture}[automata]
\node[state, initial] (0) {$q_0$};
\node[state, above right=of 0] (1) {$q_1$};
\node[state, right=of 1] (2) {$q_5$};
\node[state, below right=of 0] (3) {$q_2$};
\node[state, right=of 3] (4) {$q_6$};
\node[state, below right=of 2, secret1] (s2) {$q_8$};
\node[state, above=of s2, accepting, secret1] (s1) {$q_7$};
\node[state, below=of s2] (5) {$q_9$};
\node[state, right=of 5, accepting, secret2] (s3) {$q_{10}$};
\node[state, below=of 3, accepting] (m1) {$q_3$};
\node[state, above=of 2, accepting] (m2) {$q_4$};

\path
(0) edge node{$\sigma_0$} (1)
(1) edge node[label={[shift={(-10pt,-7pt)}]0:\faTimes}]{$\sigma_5$} (2)
(0) edge[swap] node{$\sigma_1$} (3)
(3) edge node{$\sigma_2$} (1)
(1) edge node{$\sigma_6$} (4)
(3) edge[swap] node{$\sigma_5$} (4)
(2) edge node[near end, label={[shift={(-12pt,-10pt)}]0:\faTimes}]{$\sigma_7$} (s1)
(2) edge[swap] node[label={[shift={(-10pt,10pt)}]0:\faTimes}]{$\sigma_8$} (s2)
(4) edge[swap] node[label={[shift={(-10pt,8pt)}]0:\faTimes}]{$\sigma_9$} (5)
(s1) edge node{$\sigma_8$} (s2)
(s2) edge node[label={[shift={(6pt,0pt)}]180:\faTimes}]{$\sigma_9$} (5)
(5) edge node[label={[shift={(-15pt,-7pt)}]0:\faTimes}]{$\sigma_{10}$} (s3)
(3) edge[bend left] node{$\sigma_3$} (m1)
(m1) edge[bend left] node{$\sigma_4$} (3)
(2) edge[bend left] node{$\sigma_3$} (m2)
(m2) edge[bend left] node{$\sigma_4$} (2)
;
\end{tikzpicture}
    }
    \caption{Overall control policy $\mathcal{D}$ for $\mathbf{G}$ (with protectable events converted to controllable events)}
    \label{fig:policy:example:multilevel}
\end{figure}

Finally combining the above $\mathcal{D}_1$ and $\mathcal{D}_2$ yields an overall control policy $\mathcal{D}$ as in (\ref{eq:merge:multilevel}), which is shown in \cref{fig:policy:example:multilevel}.
Observe that every string from the initial state $q_0$ that can reach the secret states in $Q_{s1}=\{q_7,q_8\}$ has at least two disabled events in $\Sigma(C_0) \cup \Sigma(C_1) \subseteq \Sigma_0 \cup \Sigma_1$. Thus the least number of protections $u=2$ and the lowest security level $v_1=0$ are satisfied. 
Moreover, every string from $q_0$ that can reach the secret state in $Q_{s2}=\{q_{10}\}$ has at least two disabled events in $(\Sigma(C_1) \cup \Sigma(C_2) \cup \Sigma(C_3)) \setminus \Sigma_0  \subseteq \Sigma_1 \cup \Sigma_2 \cup \Sigma_3$. Thus the least number of protections $u=2$ and the lowest security level $v_1=1$ are also satisfied. 

Now changing all disabled
transitions in \cref{fig:policy:example:multilevel} denoted by ``\faTimes'' to ``\faLock'', we obtain a protection policy $\mathcal{P}$ for the system ${\bf G}$ as follows: 
\begin{align*}
    \mathcal{P}(q) = \begin{dcases}
        \{\sigma_5\}, & \text{if $q = q_1$} \\
        \{\sigma_7, \sigma_8\}, & \text{if $q = q_5$} \\
        \{\sigma_9\}, & \text{if $q = q_6$} \\
        \{\sigma_9\}, & \text{if $q = q_8$} \\
        \{\sigma_{10}\}, & \text{if $q = q_9$} \\
        \emptyset, & \text{if $q \in Q \setminus \{q_1,q_5, q_6,q_8, q_9\}$}
    \end{dcases}.
\end{align*}
Finally, the minimum cost index is $i_{\min} =\max(i_{\min,1},i_{\min,2}) =3$. 

For this example, the protections of each protected event specified by the policy $\mathcal{P}$ may be implemented as follows:
\begin{itemize}
    \item $\sigma_5,\sigma_7,\sigma_8$: already described at the end of Section~4.3.
  \item $\sigma_9$: setting up the first of two-factor authentification with a security question.
  \item $\sigma_{10}$: setting up the second of two-factor authentification with a physical security key.
\end{itemize}

\section{Conclusions}\label{sec:conclusions}

We have studied a cybersecurity problem of protecting system's secrets with multiple protections and a required security level, while minimizing the associated cost due to implementation/maintenance of these protections as well as the affected system usability. Two usability-aware minimum cost secret protection problems have been formulated; the first one considers secrets of equal-importance, whereas the second considers heterogeneous secrets. In both cases, a necessary and sufficient condition that characterizes problem solvability has been derived and when the condition holds, a solution algorithm has been developed. Finally, we have demonstrated the effectiveness of our solutions with a running example.

In future work, we aim to extend the usability-aware secret protection problem to the setting of decentralized systems (which are typical in CPS), and develop efficient distributed protection policies. Other directions of extension from a broader perspective include generalizing the system model from deterministic purely-logical finite-state automaton with full observation to nondetermistic/probabilistic, timed, nonterminating, or partially-observed settings, and formulate/solve the usability-aware secret protection problem in those settings with different features.

\newpage



\bibliographystyle{elsarticle-num}
\bibliography{references}





\end{document}